\documentclass[aps,prb,showpacs,twocolumn,longbibliography, superscriptaddress]{revtex4-1}

\usepackage[dvips]{graphicx}
\usepackage{amsmath,amssymb}
\usepackage{amsfonts}
\usepackage{color}
\usepackage{placeins}
\usepackage{bbold}
\usepackage[normalem]{ulem}
\usepackage{soul}

\usepackage[usenames,dvipsnames]{xcolor}

\begin{document}

\newcommand{\Eq}[1]{Eq.~\eqref{#1}}
\newcommand{\Eqs}[1]{Eqs.~\eqref{#1}}

\newcommand{\ui}{\text{i}}
\newcommand{\diff}{\text{d}}
\newcommand{\im}{\text{Im}}
\newcommand{\re}{\text{Re}}
\newcommand{\iexp}{\text{i}}
\newcommand{\Tr}{\text{Tr}}
\newcommand{\la}{\langle}
\newcommand{\ra}{\rangle}
\newcommand{\uTHz}{\,\text{THz}}
\newcommand{\uGHz}{\,\text{GHz}}
\newcommand{\uMHz}{\,\text{MHz}}
\newcommand{\ukHz}{\,\text{kHz}}
\newcommand{\uzJ}{\,\text{zJ}}
\newcommand{\uJ}{\,\text{J}}
\newcommand{\uW}{\,\text{W}}
\newcommand{\ueV}{\,\text{eV}}
\newcommand{\uF}{\,\text{F}}
\newcommand{\ufF}{\,\text{fF}}
\newcommand{\uH}{\,\text{H}}
\newcommand{\upH}{\,\text{pH}}
\newcommand{\um}{\text{m}}
\newcommand{\uOhm}{\,\Omega}
\newcommand{\uns}{\,\text{ns}}
\newcommand{\ups}{\,\text{ps}}

\newcommand{\flq}{\frac{\Phi_0}{2\pi}}
\newcommand{\iflq}{\frac{2\pi}{\Phi_0}}
\newcommand{\wJbb}{\omega_J^{B}}
\newcommand{\IJbb}{I_c^{B}}
\newcommand{\IJaa}{I_c^{A}}
\newcommand{\IJcc}{I_c^{C}}
\newcommand{\CJbb}{C_J^{B}}
\newcommand{\CJaa}{C_J^{A}}
\newcommand{\CJcc}{C_J^{C}}
\newcommand{\IJab}{\hat I_c}
\newcommand{\CJab}{\hat C_J}
\newcommand{\wJab}{\hat \omega_J}
\newcommand{\Lhat}{\hat L}

\newcommand{\nl}{N_l}
\newcommand{\nr}{N_l+1}
\newcommand{\nI}{N_l + 1/2}

\newcommand{\phiL}{\phi_L}
\newcommand{\phiR}{\phi_R}
\newcommand{\phiBB}{\phi^{B}}
\newcommand{\phiAA}{\phi^{A}}
\newcommand{\phiCC}{\phi^{C}}
\newcommand{\phiAB}{\phi^{AB}}
\newcommand{\phiBC}{\phi^{BC}}

\newcommand{\sech}{\,\text{sech}} 
\newcommand{\tsym}{t^\ast}
\newcommand{\xsym}{x^\ast}

\newcommand{\Efl}{E_{\text{fl}}}

\title{Reversible Fluxon Logic: Topological particles allow ballistic gates along 1D paths}
\author{W.~Wustmann}
\affiliation{The Laboratory for Physical Sciences, College Park, MD 20740, USA}

\author{K.D.~Osborn}
\email{osborn@lps.umd.edu}
\affiliation{The Laboratory for Physical Sciences, College Park, MD 20740, USA}
\affiliation{The Joint Quantum Institute, University of Maryland, College Park, MD 20742, USA}

\begin{abstract}
As we reach the end of Moore's law, digital logic uses irreversible logic gates whose energy consumption has been scaled toward a lower limit.
Reversible logic gates can provide a dramatic energy-efficient alternative, but rely on reversible dynamics. 
Here we introduce a set of superconducting reversible gates that are powered alone by the inertia of the digital input states, contrasting existing adiabatic prototypes which are powered by an external adiabatic drive. 
The classic model of an inertia-powered reversible gate uses ballistic particles which scatter in 2D, where the digital state is represented by the particle path. 
Our ballistic gates use as the bit state the topological charge (polarity) of a fluxon moving along a Long Josephson Junction (LJJ) such that the particle path is confined	 to 1D.  
The fundamental structures of our Reversible Fluxon Logic (RFL) are 1-bit gates which consist of two LJJs connected by a circuit interface that comprises three large-capacitor Josephson junctions (JJs). 
Numerical simulations show how a fluxon approaching the interface under its own inertia converts its energy to an oscillating evanescent field, from which in turn a new fluxon is generated in the other LJJ. 
We find that this resonant forward-scattering of a fluxon across the interface requires large capacitances of the interface JJs because they enable a conversion between bound-evanescent and traveling fluxon states (without external power). 
Importantly, depending on the circuit parameters, the new fluxon may have either the original or the inverted polarity, and these two processes constitute the fundamental Identity and NOT operations of the logic. 
Based on these 1-bit RFL gates, we design and study a related 2-bit RFL gate which shows that fluxons can exhibit conditional polarity change.  
Energy efficiency is accomplished because only a small fraction of the fluxon energy is transferred to modes other than the intended fluxon. 
Simulations show that over $97\%$ of the total fluxon energy is preserved during gate operations, in contrast to irreversible gates where the entire bit energy is consumed in bit switching. To provide insight into these phenomena, we analyze the 1-bit gate circuits with a collective-coordinate model which describes the field in each LJJ as a combination of fluxon and mirror antifluxon. 
This allows us to reduce the many-junction circuit (the 3 interface JJs and the many JJs approximating the LJJs, solved numerically) to that of two coupled degrees of freedom that each represent a particle.
The evolution of the reduced model agrees quantitatively with the full circuit simulations and validates the use of the mirror-fluxon ansatz. 
Parameter tolerances are calculated for the proposed circuits and indicate that RFL gates can be manufactured and tested. 
\end{abstract}

\maketitle

\section{Introduction}

Today's conventional digital logic is based on irreversible gates. 
Physically, the irreversibility arises in a gate due to the energy that is dissipated in switching processes between the states representing the logic bits. 
This energy cost is generally set by the energy of the digital state itself: the charging energy of a voltage state in semiconductor logic or the magnetic energy of a flux state in conventional superconducting logic. 
In both cases, sufficiently large damping ensures deterministic and fast state switching compatible with GHz processing speeds. 
Unlike CMOS gates, however, the intrinsic damping in superconducting circuits is very small, allowing nearly dissipationless reversible dynamics. 
As a result, superconducting circuits also allow for the implementation of both quantum logic and reversible digital logic. 
The most common type of reversible digital logic is ``adiabatic reversible logic'', where state switching uses an external waveform (or clock) to steer an adiabatic state evolution in the absence of large damping. 
Here we report on an unusual gate class known as ballistic reversible gates, which are powered by the inertia of the incoming bits. 
Specifically, we find that fluxons can undergo energy conserving transformations of polarity and this mechanism can be used for ballistic reversible logic gates.

Industrial development of microelectronics has long benefitted from transistor density scaling which allows the vast improvements described by Moore's law. 
However, the related performance scaling has slowed down considerably in recent years and is expected to come to an end soon. 
One particular limiting factor is the heat generated during CMOS transistor switching in logic gates \cite{IRDS2017, TheWon2017}. 
These logic gates are irreversible from an information perspective because they do not perform one-to-one maps of input-to-output states. 
Their energy cost includes Landauer's entropy cost of $\ln(2) k_B T$ for the erasure of each bit of information at temperature $T$. 
Though the bit switching energy of irreversible gates could in principle be this small, in practice it consumes the entire energy of the bit state itself (e.g., the voltage state) through fast damping in order to enable GHz processing speeds. 
For reasons that include thermal stability and state distinction, the bit state energy has to be $\gg k_B T$ (e.g., $> 1000 k_B T$), and consequently the switching energy exceeds Landauer's entropy cost by orders of magnitude in practice. 
Similar principles hold for conventional single-flux quantum (SFQ) logic in superconducting circuits. 
SFQ logic uses a flux quantum generated by a persistent current in a quantizing loop to represent one bit state. 
Bit switching, i.e. the change of the flux state in the loop, is enabled by a Josephson junction (JJ); 
once the total current acting on a JJ exceeds its critical current $I_c$, it undergoes a rapid $2\pi$-phase change. 
Equilibrium is reestablished through a damping resistor that shunts the JJ \cite{LikSem1991}. 
Similar as in the CMOS bit switching, the dissipated energy here is on the order and larger than the bit energy of the flux quantum itself; specifically, the switching energy is $\sim I_c \Phi_0$,  where $\Phi_0$ is the superconducting flux quantum. One such recently developed (irreversible) SFQ logic type \cite{HerrETAL2011, PrivComm2017, Holmes2013} allows $2\pi$-phase switching at an energy of only $I_c \Phi_0 \approx 1300 k_B T$.

Reversible logic gates provide an alternative approach to computing. 
These gates produce no entropy related to bit-erasure, since they perform one-to-one mapping of input-to-output states. 
A corresponding theoretical model for a reversible computer was established decades ago \cite{Bennett1973}. 
To provide an advantage over irreversible logic, reversible gates must use dynamics that conserves most of the bit-state energy.
This can be achieved within an adiabatic model, developed by Likharev \cite{Lik1982} in 1982, by using externally applied fields that adiabatically modulate the circuit potential. 
Arriving later than their quantum counterparts in superconductivity, reversible digital logic has been demonstrated more recently in circuits, which include N-SQUID \cite{RenSem2011} and reversible QFP \cite{QFPgate}. 
According to Likharev's adiabatic model, energy dissipation varies proportional to the modulation frequency \cite{Lik1982, YoshikawaETAL2015} and thus can be made
arbitrarily small as the speed is lowered.

Though demonstrated reversible gates are of the adiabatic (reversible) logic type, a physically distinct type known as ballistic gates can be studied. 
These make use of scattering processes and are solely powered by the initial energy of the digital state. 
This is apparent in the original ``billiard ball model'' \cite{FredTof1982}: particles moving under their inertia collide with each other and with reflective boundaries of a well-defined 2D gate geometry. 
The resulting paths taken by the particles encode the digital output state of the gate. 
Ballistic gates have been studied using soliton propagation within fibers \cite{IslSoc1991} or 2D layered media \cite{SchOre2005}. 
A technical challenge in this type of reversible logic is that they must be correctable for path perturbations in 2D and desynchronization errors \cite{DruMan1994}, and here we will address the former.

In this work we introduce gates for efficient ballistic reversible logic in superconducting circuits under the name of Reversible Fluxon Logic (RFL). 
This approach uses fluxons and antifluxons in Long Josephson junctions (LJJs) to represent the two bit states. 
A fluxon -- or flux-soliton -- is a spatially extended topological excitation of the LJJ and carries a quantized magnetic flux $\Phi_0$ ($-\Phi_0$ in case of an antifluxon). 
In RFL the LJJs form fluxon-waveguides and are connected at circuit-interfaces. 
Fluxon gates can be made of these structures where a fluxon scatters from one LJJ to another through a resonant nonlinear process.
Through numerical simulation of the gate circuits we find cases where an incoming fluxon exites resonant dynamics of the coupled LJJs  
which results in the net forward scattering of the fluxon to another LJJ.
In these nonlinear processes the character of the incoming fluxon changes near the interface where it turns into a localized interface excitation with evanescent fields in the LJJs.  
From this localized state a new ballistic fluxon forms afterwards in the other LJJ.
For this 1D scattering no path corrections are required in contrast to the original (2D) ballistic gate model.

Importantly, while the fluxon number is conserved, the scattering can result in the transformation of a fluxon into an antifluxon (or vice versa), and this change of polarity provides a means for bit-switching (defined as a NOT gate). 
This is fundamental because the fluxon polarity represents a topological charge which cannot change in a planar LJJ (only along a LJJ that is twisted in 3D). 
The NOT gate occurs in a 1-bit structure which has three JJs in the interface. 
For different parameters of these interface JJs the dynamics changes; in particular the gate type can be changed to an Identity (ID) gate, where fluxon polarity is preserved but the resonance is changed. 
In related 2-bit gates the fluxon polarities induce a conditional polarity change. 
The bit-switching in RFL involves a (gradual) undamped $4\pi$-phase change of an interface JJ. 
The $4\pi$ change is thus found to be enabled by resonance between LJJs, and it contrasts adiabatic reversible logic which generally uses a $\sim 2\pi$ phase change, related to neighboring minima of the JJ potential \cite{Lik1982}.

The RFL gate interface is designed to ensure that an incoming fluxon (with a given velocity and for given LJJs) is coherently transformed into the local interface excitation and back into another forward-moving fluxon in a different LJJ. 
To accomplish this, the ends of the LJJs near the interface must behave differently than in bulk. 
In particular we find that relatively large (added shunt) capacitances of the interface JJs are crucial to the resonant scattering present in the 1- and 2-bit gates.
We have designed the RFL gates for relatively high incoming fluxon velocities of around $0.6 c$, where $c$ is the upper (relativistic) velocity limit. 

The RFL gates presented in this study are reversible in the sense that their energy cost is only a small fraction of the digital state energy. 
In our simulations an output fluxon can obtain $97 \%$ of the energy of the input fluxon.
The gates can also be logically reversible at lower efficiency, which allows some tolerance to imperfect structures.
The remaining small fraction of initial energy is dissipated to the environment of the gate in the form of small-amplitude plasma waves in the LJJs. 
The JJs in the simulations are undamped. Beyond this general property of superconducting circuits, we note that the observed high energy retention in the fluxon degree of freedom is a speciality of our resonant logic gates. 
Also, the RFL gates have a gate time of only a few Josephson oscillation periods, $T_{\text{gate}} \sim 1/\nu_J$, where $\nu_J$ is the natural frequency in the LJJ. 
In principle $\nu_J$ could be on the order of tens of GHz for fast processing speeds.

The complex scattering dynamics observed at the circuit interfaces in an RFL gate goes beyond the usual fluxon perturbation theory \cite{McLaughlinScott1978}. 
While the latter assumes the fluxon's integrity (possibly allowing for internal excitations, shape modes, etc.) in our case the fluxon breaks up at the interface and its energy is converted into an interface oscillation mode. 
This mode involves an excitation of the interface JJs but also has finite amplitude in the adjacent LJJs, decaying away from the interface. 
The LJJs therefore play a role not only as input and output ports, but also as integral parts of the gate itself.
To analyze the non-perturbative scattering dynamics we parametrize the fields in each LJJ as a superposition of a fluxon paired with a mirror antifluxon. This collective coordinate (CC) model can account for the various scattering processes with conserved energy, including those where the fluxon polarity changes. In case of the 1-bit RFL gates, for example, the Lagrangian produces coupled dynamics in 2 coordinates. The dynamics can thus be thought of as a possible excitation in one LJJ interacting with a possible excitation in the other LJJ, where each has an independent spatial coordinate along its LJJ. The CC dynamics accurately reproduces dynamics from the full gate simulation. The forward-scattering with and without polarity change are seen to arise from the special form of the effective potential and essential mass-gradient forces generated by the interface. The mass-gradient forces, not used in other SFQ logic, are introduced in our superconducting digital gates through engineered capacitances. 

The interface scattering of the fluxon may be compared with scattering of a fluxon at a point defect in an LJJ \cite{FeiKivVaz1992, GooHab2007}, or at a qubit-generated perturbation potential \cite{AveRabSem2006, UstinovETAL2014, AnnaFluxon, KuzminETAL2015}. 
In those situations the moving fluxon interacts with a small-amplitude bound state at the defect. 
Above a critical velocity the fluxon is transmitted, but a slow fluxon may be back-scattered or trapped. Unlike in these systems, our system allows polarity change of a fluxon. 
This is here made possible since one of the superconducting electrodes of the LJJ is interrupted in the interface by a JJ. 
This JJ can undergo large phase winding in resonance, corresponding to a change of the flux inside the LJJ, e.g. by two flux quanta in case of the NOT gate. 
Moreover, we typically find that the outcome of scattering at the interface depends only moderately on the incoming velocity, in contrast to fluxon scattering at point defects for which high sensitivity to the incoming fluxon velocity had been found \cite{FeiKivVaz1992}. 

Building on the results for the 1-bit gates, we have also designed and simulated  2-bit gate structures which are found to allow conditional polarity changes for the two incoming fluxons. In this particular 2-bit gate, a NSWAP, the coupled dynamics of the two input fluxons is symmetry-related to the dynamics of two different uncoupled 1-bit gates. Depending on the relative polarity of the two input fluxons it conditionally produces a NOT or ID operation. 
This 2-bit gate has an interface with 7 capacitance-shunted JJs. 
The present work covers an initial set of RFL gates. Subsequent work describes how to store and launch fluxons for the purpose of synchronization and resupply of energy between gates \cite{WusOsb2018}.  
Ref.~\cite{WusOsb2018} also uses a 2-bit gate named IDSN with similar dynamics as in the here presented gates, as a key component of an efficient CNOT gate.

The paper is organized as follows:
In Sec.~\ref{sec:operation} we introduce the gate circuit and briefly describe the main results for the fluxon logic gates. 
We start by presenting numerical simulations of circuits with fluxon dynamics in the 1-bit gate structures. 
We then analyze the fluxon dynamics in these gate structures by means of a collective coordinate (CC) model. 
It describes the fields near the interface as consisting of a fluxon paired with a mirror antifluxon (Sec.~\ref{sec:operation_CCM}). 
From two 1-bit gate types we then construct a 2-bit gate which likewise is unpowered and reversible (Sec.~\ref{sec:operation_2bit}). 
We further study dynamics of the 2-bit gate embedded into a simulation test platform for fluxon launch and output state storage. 
This serves to show that the gates do not depend sensitively on a perfectly launched fluxon.
Section~\ref{sec:CCM} presents more detail on the CC model and its application to describe the fluxon scattering at the circuit interfaces, including such scattering that is not used for gates. 
This analysis provides a better intuition for the scattering dynamics and helps to identify the relevant parameter regimes for the gates. 
The precise interface parameters used in gates are calculated numerically in Sec.~\ref{sec:optimization} from a Monte-Carlo optimization of parameters. 
The gate operation under changes of single parameters is calculated as well (parameter margins).
In Sec.~\ref{sec:2bitgates} we explain in detail the operation of the 2-bit gate and how it can be mapped to equivalent 1-bit gates, as it depends on the relative polarity of input fluxons. 
A conclusion is given in Sec.~\ref{sec:summary}. 
The appendix contains details of the gate analysis and shows that we expect negligible energy loss from fluxons in our LJJs due to the small LJJ-model discreteness.

\begin{figure*}[tb]
\includegraphics[width=\textwidth]{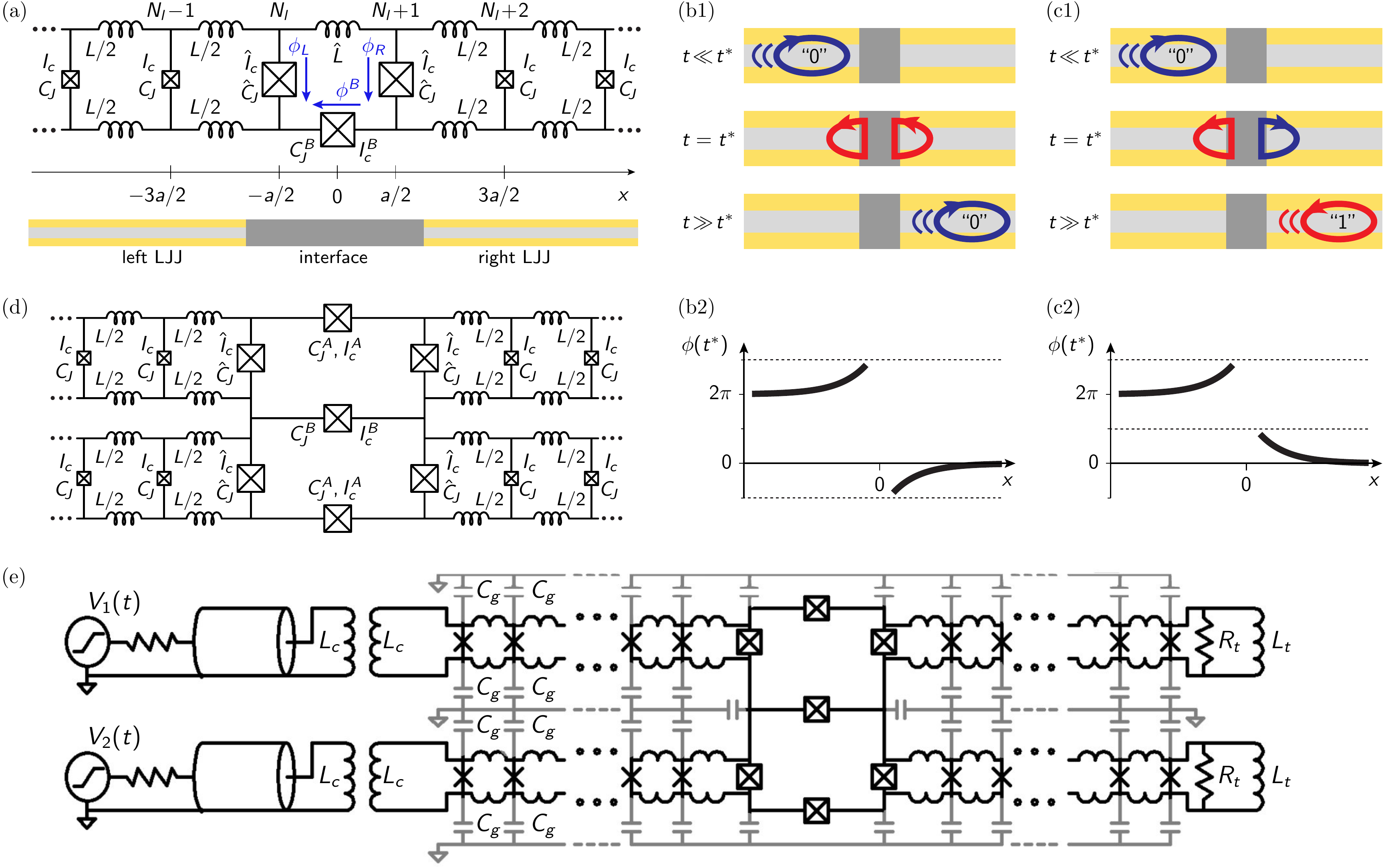}
\caption{
(a) 
General 1-bit gate circuit composed of two LJJs ($n\leq N_l-1$ and $\geq N_l+2$) and an interface that contains JJs with phases $\phiL$, $\phiR$, and $\phiBB$. 
Depending on the interface parameters (capacitances and critical currents of interface JJs, inductance) the structure allows various types of fluxon dynamics, including the resonant forward scattering to be employed in RFL gates. 
As indicated in the diagram below, the LJJs with cell inductance $L$ and critical current $I_c$ are  discrete versions of continuous LJJs (yellow and light gray JJ trilayer boxes) with inductance and critical current per unit length, $L/a$ and $I_c/a$ and $a \to 0$.
(b,c) 
An illustration of 1-bit gate phenomena found in numerical solutions, using a free traveling fluxon as an initial condition (initially several Josephson penetration depths away from the interface). 
Fluxons and antifluxons encode the bit states ``0'' and ``1''. 
Currents are shown at the scattering symmetry time $\tsym$, as well as before and after, for the ID gate (b1) and the NOT gate (c1). 
Panels (b2) and (c2) show the LJJ-phase profiles at $\tsym$, 
indicating the then excited localized state with evanescent fields into the bulk of the LJJs.
The corresponding numerical solutions are shown in Fig.~\ref{fig:4bounce} and \ref{fig:2bounce}, and solutions of an analytical model in Fig.~\ref{fig:CCM}.
(d) 
2-bit gate circuit consisting of two in- and two output LJJs connected by an interface.
Fig.~\ref{fig:2bit_gate} shows the simulated operation of a specific 2-bit gate (NSWAP).
(e)
2-bit gate circuit (central part) in a simulation test platform. 
The platform allows simulations of gates with non-ideal fluxon launch: voltage steps created in the transmission lines launch fluxons into the inductively coupled LJJs. 
A ground plane (grey wires) adds stray capacitance. The fluxons move towards the gate where they interact. A successful gate will result in forward-scattered fluxons, which then induce circulating currents in the storage loop (right) for possible readout.
The simulated operation of a specific 2-bit gate (NSWAP) in the test platform is shown in Fig.~\ref{fig:2bit_gate_testplatform}. 
}
\label{fig:interface}
\end{figure*}

\section{Operation and analysis of gates}
\label{sec:operation}

\subsection{1-bit gate system}

We study the dynamics of a fluxon in a superconducting circuit as sketched in Fig.~\ref{fig:interface}(a). It consists of two discrete LJJs
at $|x| \geq a/2$. These are made from an array of Josephson junctions (JJs), 
with parallel critical current $I_c$ and capacitance $C_J$, 
connected to their neighbors through cell inductance $L$.
Each LJJ ends at an interface Josephson junction of capacitance and critical current  ($\CJab, \IJab$). 
The termination junctions ($\CJab$, $\IJab$) of two LJJs are connected in a series loop with two other elements: a central interface junction of ($\CJbb$, $\IJbb$), 
and an inductor $\Lhat$. 
Elements ($\CJbb$, $\IJbb$), ($\CJab$, $\IJab$) and $\Lhat$ with their connections constitute the interface cell.

We simulate the dynamics by numerically integrating the $N+1$ equations of motion 
for the junction phase differences: $\phiBB$ for the center interface junction, 
and $\phi_n$ for the $1 \leq n \leq N_l$ and $N_l+1 \leq n \leq N$
junctions in the left and right half of the circuit, respectively, 
where the phases of the left and right interface junctions are 
$\phiL = \phi_{\nl}$ and $\phiR = \phi_{\nr}$. 
The equations of motion are generated by the Lagrangian
\begin{eqnarray}
\label{eq:Lagr_orig_discrete}
 \mathcal{L}
&=& \left(\flq\right)^2 \left[ \sum_n \frac{C_{J,n}}{2} (\dot{\phi}_n)^2 
+ \frac{\CJbb}{2} (\dot{\phi}^{B})^2 \right] \\
&-& \left(\flq\right) \left[\sum_n I_{c,n} ( 1 - \cos\phi_n )  
+ \IJbb ( 1 - \cos\phiBB ) \right]  \nonumber \\
&-& \frac{1}{2} \sum_n \left[ L_n^A (I_n^A)^2  +  L_n^B (I_n^B)^2 \right]
 \nonumber 
\end{eqnarray}
As defined above, there are identical junctions ($C_{J,n} = C_J$, $I_{c,n} = I_c$) 
and inductors, $L_n = L$, in the LJJ cells.
Also, the left and right interface junctions have
$C_{J,\nl} = C_{J,\nr} = \CJab$ and 
$I_{c,\nl} = I_{c,\nr} = \IJab$.
Finally, the interface inductance and center junction have values $L_{\nl-1} = \Lhat$ and 
($\CJbb$, $\IJbb$), respectively.
The currents on the upper rail of the LJJs are given by
$I_n^A = \Phi_0 \left(\phi_{n+1} - \phi_{n}\right)/(2\pi L_n)$
($n<\nl$, $n \geq \nr$)
and  $I_n^B = - I_n^A$ on the lower, where $L_n = L^A_n + L^B_n$
is the total inductance of the $n$--th LJJ cell.
The current on the upper rail in the interface (with center inductor $\Lhat$) is
given by
$I_{\nl}^A = \Phi_0 \left(\phiR - \phiL + \phiBB \right)/(2\pi \Lhat) = -I_{\nl}^{B}	$, where $I_{\nl}^{B}$ is the current on the lower rail of the interface.
The interface cell is symmetric in propagation direction (left--right),
as required for a physically reversible gate.

When scaling the parameters $(L, C_J, I_c) \propto (a, a, 1/a)$, 
in the limit $a \to 0$ a continuous LJJ is realized 
rather than our approximate one. The dynamics of the continuous LJJ are governed by the 
sine-Gordon equation (SGE) \cite{DauxoisPeyrard__PhysicsOfSolitons, McLaughlinScott1978, Newell},
\begin{equation}\label{eq:SGE}
 \ddot \phi - c^2 \phi'' + \omega_J^2 \sin\phi = 0
 \,.
\end{equation}
The characteristic time and length scales of the LJJ are given by the
Josephson frequency $\omega_J$ and Josephson penetration depth $\lambda_J$,
defined by $\omega_J^2 = (2\pi/\Phi_0) I_c C_{J}^{-1}$
and $\lambda_J^2 = (\Phi_0/2\pi) a^2 (L I_c)^{-1}$,
where
$L/a$ and $I_c/a$ are the LJJ inductance and critical current per unit length $a$.
The upper bound for the group velocity
in the LJJ is the Swihart velocity $c = \lambda_J \omega_J$.

In the (infinite) continuous LJJ a stable fluxon exists in form of the 
soliton solution to the SGE, with the phase and phase-derivative profiles
\begin{eqnarray}\label{eq:soliton}
 \phi^{(\sigma, X)}(x,t) 
 &=& 4 \arctan \exp\left(-\sigma \left(x - X(t)\right) \big/ W \right) \\
 \dot \phi^{(\sigma, X)}(x,t) 
 &=& \frac{2\sigma \dot X}{W}\sech\left(\left(x - X(t)\right) \big/ W \right)
\,,
\end{eqnarray}
where $\dot \phi (\Phi_0/2\pi)$ is the voltage across the LJJ.
We choose the range $0 \leq \phi^{(\sigma, X)} \leq 2\pi$, and the polarity 
$\sigma = 1$ ($\sigma = -1$) for a fluxon (antifluxon) solution.
Here $X$ is the center position of the fluxon, 
which propagates with constant velocity $v_0 = \dot X$ ($|v_0| < c$) and 
has a characteristic width $W = \lambda_J \sqrt{1 - v_0^2/c^2}$.

For use in a logic gate, a fluxon first needs to be created in an LJJ.
In this context, Fig.~\ref{fig:interface}(e) shows a schematic of a simple test platform for a 2-bit gate, with two input and two output LJJs 
(the 2-bit gate is discussed in Sec.~\ref{sec:operation_2bit}, and in more detail in Sec.~\ref{sec:2bitgates}).
In addition to the 2-bit gate itself, the platform provides components that allow one to 
(i) create ballistic fluxons in the input LJJs from coupled transmission lines and 
(ii) store flux at the end of each LJJ output for a gate readout (test).
The fluxon launch in each input LJJ is initiated by ramping up the input voltage of a transmission line. Also, in the platform the LJJs and interface are additionally coupled via small capacitances to ground.
We have successfully simulated the fluxon launch and the subsequent gate operations and flux storage of this test platform, 
as discussed in Sec.~\ref{sec:operation_2bit}, cf.~Fig.~\ref{fig:2bit_gate_testplatform}. 
In the interest of simplicity, simulations presented here are mainly performed only for the gate alone, 
without circuit components for launch, output flux storage, and capacitive coupling to ground. 

In these simulations, a soliton is simply taken as the initial condition, 
$\phi_n(t=0) = \phi^{(\sigma, X)}(x_n,0)$. 
This is evaluated on the discrete lattice of the circuit,
$x_n = a n - a\left(\nI\right)$ ($n=1\ldots N$),
where $a$ is the lattice spacing. 
Specifically, a fluxon is initialized in the left LJJ, $X(t=0) = X_0 < 0$,
and is incident on the interface with initial velocity $\dot X = v_0 > 0$,
and polarity $\sigma=1$.

In an ideal continuous LJJ the fluxon energy,
$E_{\text{fl}} = 8 E_0/\sqrt{1 - v_0^2/c^2}$, is conserved according to the SGE,
where $E_0 = \Phi_0 I_c \lambda_J/(2\pi a)$.
In a discrete LJJ the fluxon motion is in general damped, 
because the discreteness forms a perturbation through which a moving fluxon can excite linear plasma modes of the (discrete) LJJ. (These wave modes are solutions of the linearized SGE.) 
The moving fluxon thus emits plasma waves and as a result loses energy \cite{BraunKivshar1998, UstinovETAL2008, PeyKru1984}.
However, this energy loss mechanism is strongly suppressed if $a/\lambda_J<1$, 
and this criterium characterizes the regime of `small LJJ discreteness'. 
The LJJs used in this paper as part of the logic gates are chosen in this regime, 
$(a/\lambda_J)^2 = 2 \pi I_c L/\Phi_0 = 1/7$,
where the energy loss of the moving fluxon is negligible. 
In App.~\ref{app:discreteness} we briefly discuss the fluxon-energy loss-rate at this (small) discreteness and its dependence on the initial fluxon velocity. 
For a typical initial velocity, $v_0/c=0.6$, the fluxon energy is 
$E_{\text{fl}} = 10 E_0$, where $20 \%$ of $E_{\text{fl}}$ is kinetic energy. 
As shown in App.~\ref{app:discreteness}, in our LJJs such a fluxon loses only a fraction $10^{-7}$ of its energy in a time $\omega_J^{-1}$ due to the discreteness.
This allows us to discuss the LJJs as nearly continous in the context of the RFL gates, which are meant to operate over such short times.

\subsection{Fluxon forward scattering for 1-bit gates}

The gate circuit of Fig.~\ref{fig:interface}(a) supports equilibrium states of the form 
$\phi(x) = 2\pi K_L \Theta(-x) + 2\pi K_R \Theta(x)$ 
in the LJJ and $\phiBB = 2\pi (K_L - K_R)$ on the center interface JJ, 
where $K_{L,R} \in \mathbb{Z}$ and $\Theta(x)$ is the Heaviside step function. 
Since we constrain studies to a parameter regime with $L < \Phi_0/(2\pi \IJbb)$, 
without loss of generality we can disregard states with finite flux trapped in the interface cell \cite{Barone},
such that $\phiL-\phiR + \phiBB = 0$ in equilibrium.
Taking $(K_L, K_R)=(0,0)$ as the initial state, 
in fully inelastic scattering of an input fluxon, i.e.~if the entire initial fluxon energy is exhausted in the interface region,
the interface would settle into the equilibrium state $(K_L, K_R)=(1,0)$.
Inelastic scattering generally results from excitation of
high-frequency plasma waves ($\omega > \omega_J$) at the interface
which are radiated into the LJJs and thus spread the initial fluxon energy incoherently.
The amount of radiation generated at the interface depends on the interface parameters.
It is strongly suppressed in the following regimes, 
related to known scattering dynamics in an LJJ:
(i)
The fluxon is transmitted across the interface
for $\IJbb \gg I_c$ and $\CJbb \ll \IJbb C_J/I_c$,
while $\IJab \simeq I_c$, $\CJab \simeq C_J$, $\Lhat \simeq L$, 
because the center interface junction essentially acts as a small linear inductance due to the large $\IJbb$, and the entire circuit thus approximates a single LJJ.
(ii)
If the potential energy of the interface is too high, e.g.~$\IJab \gg I_c$,
the fluxon is reflected back before reaching the interface, similar to a shunt-terminated LJJ.
(iii)
The inter-LJJ coupling is suppressed due to the small $\CJbb$ and $\IJbb$
if 
$\CJbb \ll C_J$, 
$\IJbb \ll I_c$,
$\IJab \simeq I_c$, $\CJab \simeq C_J$ and $\Lhat \leq L$.
This is comparable to an open terminated LJJ and thus
the incident fluxon is scattered back elastically as an antifluxon,
while the center interface junction undergoes a $4\pi$-phase winding.
(Note that the interface cell does not store significant flux, 
but 2 flux quanta, $2 \Phi_0$, must be exchanged between the interface cell and the environment.)
These elastic processes involve transitions 
of the initial equilibrium state $(K_{L}, K_{R}) = (0,0)$ in the interface region to 
(i) $(K_{L}, K_{R}) = (1,1)$,
(ii) $(K_{L},  K_{R}) = (0,0)$, and
(iii) $(K_{L},  K_{R}) = (2,0)$, respectively.

Inelastic scattering and fluxon reflection are undesirable for efficient reversible gates. 
We therefore mainly report on reversible scattering phenomena
where the fluxon energy is transferred into a coherent localized oscillation
about the equilibrium state $(K_L,K_R)=(1,0)$, 
and remarkably followed by the creation of a fluxon or antifluxon in the other LJJ
as a resonant forward-scattering process.

\begin{figure}[tb]
\includegraphics[width=\columnwidth]{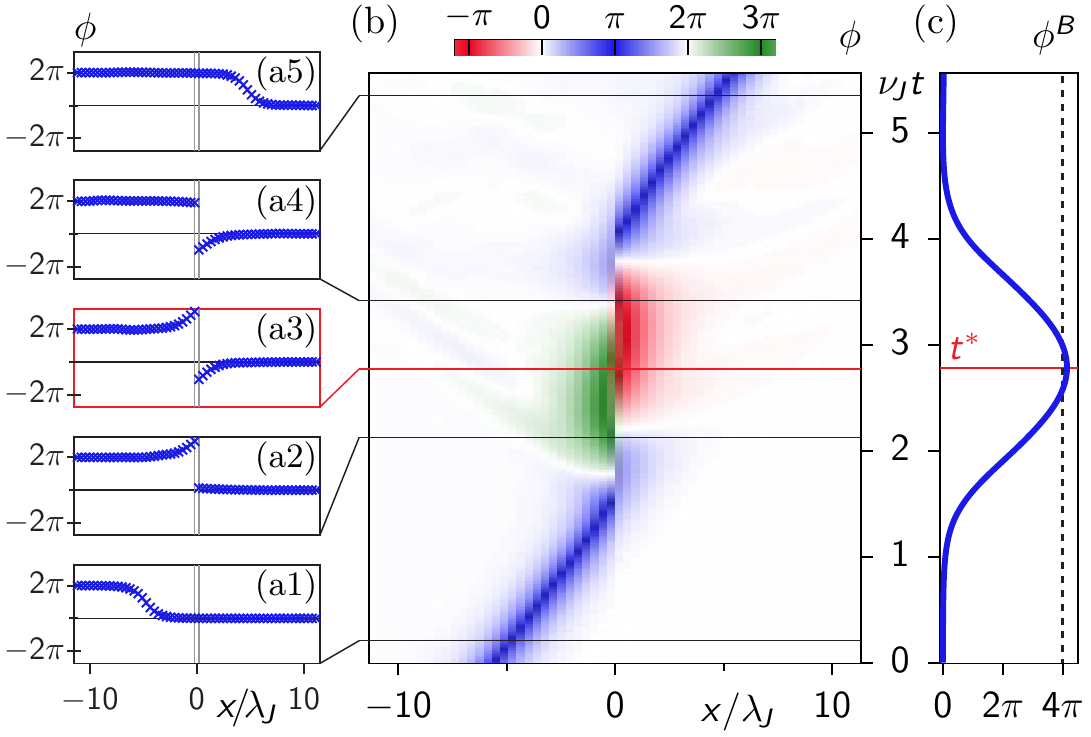}
\caption{
Polarity-preserving ID gate realized with 1-bit gate structure, Fig.~\ref{fig:interface}(a):
(a) LJJ phases $\phi_n$ vs.~$x_n$ at specific times, $|t-\tsym| \sim 1/\nu_J$,
where $\tsym$ is gate symmetry time and $\nu_J$ is the Josephson frequency. 
(b) color representation of $\phi_n$
vs.~$x_n$ and $t$,
and
(c) phase $\phiBB$ of center interface junction,
for fluxon incident on interface at $x=0$
with initial velocity $v_0 = 0.6 c$.
In panel (b) the blue color of the fluxon trace at $x<0$ and $x>0$ indicates the same fluxon type. 
The interface parameters are
$\CJbb/C_J = 6.0$, $\IJbb/I_c = 0.10$, $\CJab/C_J = 5.8$, $\IJab/I_c = 0.80$, and $\Lhat/L = 0.06$,
and we note that the center junction has negligible critical current ($\IJbb$) but significant capacitance ($\CJbb$). 
}
\label{fig:4bounce}
\end{figure}
In Fig.~\ref{fig:4bounce}(b) numerically simulated dynamics of the LJJ phases $\phi_n(t)$ are shown vs.~position $x_n$ and time $t$, for interface parameters given in the caption. 
The phases $\phi_n$ are also shown for specific times $t$ in Fig.~\ref{fig:4bounce}(a1-a5). 
After moving ballistically (approximately freely) in the left LJJ, e.g. panel (a1), 
the fluxon energy is converted to a localized oscillation at the interface with evanescent fields into the LJJs, e.g. panels (a2-a4).
During that stage the characteristic phase-profile of the original fluxon, \Eq{eq:soliton}, is destroyed. 
A large phase difference $\phiL-\phiR$ has accumulated 
between the left and right interface junctions, 
$\phiL=\phi_{N_l}$ and $\phiR=\phi_{N_l+1}$.
This is accompanied with an increase of the phase $\phiBB$ 
of the center interface junction, shown in Fig.~\ref{fig:4bounce}(c),
such that the phase across $\Lhat$ remains small,
$(\phiR-\phiL+\phiBB) \ll \pi$.
During the localized oscillation, the field $\phi(x,t-\tsym) - 2\pi \Theta(-x)$,
measured relative to the equilibrium state $(K_L,K_R)=(1,0)$,
behaves parity-time antisymmetric 
where the symmetry time $\tsym$ is defined by $\dot{\phi}^B(\tsym) = 0$.
After the symmetry time $\tsym$ we observe the energy from the local oscillation 
create a ballistic (unpowered) moving fluxon in the right LJJ (a5), 
while the interface region is left in the equilibrium state $K_L=K_R=1$.
The process is almost ideally elastic, 
with the new fluxon propagating at $96 \%$ of the initial velocity $v_0$,
and we note that no bias is present in the gate region (within several Josephson penetration depths of the interface).
Compared to straight transmission across the interface the new fluxon 
appears with a time-delay of $T \approx 2.7 \cdot (2\pi/ \omega_J)$, 
within a factor of three of the Josephson period.

\begin{figure}[tb]
\includegraphics[width=\columnwidth]{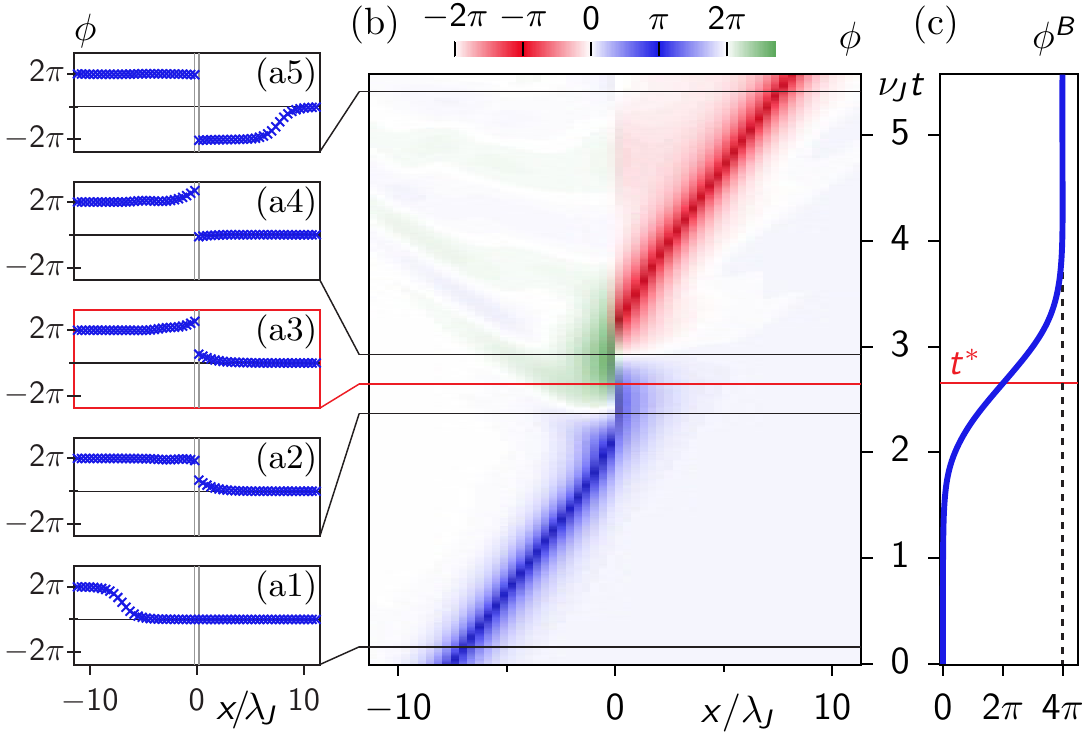}
\caption{
Polarity-inversion (NOT) gate realized with 1-bit gate structure.
Subfigures and parameters as in Fig.~\ref{fig:4bounce}, 
except for $\CJbb/C_J = 12.0$.
In panel (b) the blue color of the fluxon trace at $x<0$ and red color at $x>0$
indicate a conversion from fluxon to antifluxon. Even though the center junction has 
wound by $4\pi$, the interface stores no flux before or after the gate. 
}
\label{fig:2bounce}
\end{figure}

A different reversible forward-scattering 
is illustrated in Fig.~\ref{fig:2bounce},
for an interface that differs from that in Fig.~\ref{fig:4bounce} only by 
increasing $\CJbb$ to $12.0 C_J$.
Coherent energy transfer again takes place from the incident fluxon to 
an interface oscillation which spans a time $T \approx 1.1 \cdot (2\pi/ \omega_J)$, 
approximately half of the process in Fig.~\ref{fig:4bounce}.
The field $\phi(x,t-\tsym) - 2\pi\Theta(-x)$ 
exhibits parity-time symmetry with respect to a time $\tsym$ defined by $\phiBB = 2\pi$.
Thus, after the oscillation the interface region settles to the state with $K_L=1$, $K_R=-1$. 
Here a fluxon is emitted into the right LJJ, (a5), 
but in contrast to Fig.~\ref{fig:4bounce} it is 
an \emph{anti}fluxon (a fluxon with inverted polarity, $\sigma=-1$).

If the two fluxon polarities $\sigma = \pm 1$ encode the bit states 0 and 1,
these unconventional fluxon scattering phenomena implement 1-bit reversible gates. 
The process of Fig.~\ref{fig:4bounce} (Fig.~\ref{fig:2bounce}) performs an ID (NOT) operation by preserving (inverting) the polarity of the input fluxon,
during which only $2.1$\% ($2.6$\%) of the initial fluxon energy 
$E_{\text{fl}} = 10 E_0$ is dissipated.  
This results in a dissipation $< 0.03 E_{\text{fl}} \ll U_{\text{fl}} = 8 E_0$, much less than the potential energy (rest energy) of the fluxon $U_{\text{fl}}$.

The small energy difference between the input fluxon and the output fluxon
is dissipated in form of plasma waves generated at the interface and radiated into the LJJs, see the faint wave patterns in Figs.~\ref{fig:4bounce} and \ref{fig:2bounce}. 
In addition to this loss mechanism we expect smaller losses from (superconductor) quasiparticles and dielectric loss; methods to control these mechanisms are known, 
even in qubits where the sensitivity is much higher to loss than our gates.

The interface parameters of Figs.~\ref{fig:4bounce} and ~\ref{fig:2bounce} were chosen near numerically optimized values for the two respective gate types. 
The optimization studies are presented in Sec.~\ref{sec:optimization}.
As described above, a conversion between the two gate types is here accomplished
by adjusting the capacitance $\CJbb$ alone.
Another parameter change has a similar effect, namely an increased $\IJbb$
can turn an ID into a NOT 
(compare Fig.~\ref{fig:CCM_Epot_others}(c2)).
It is important to emphasize that each gate by itself is fully autonomous, 
requiring no drive fields, and is determined by the fixed circuit parameters alone. 
The efficient gate dynamics takes place 
on the time scale of the inverse plasma frequency $2\pi/\omega_J$.

These 1-bit gates operate with one fluxon (bit) at a time.
Therefore the spacing required between fluxons must be greater than the sum of two terms.
The first term is to avoid perturbations of the gate dynamics by a second arriving fluxon, and is given by the gate time multiplied by the velocity.
The second term is related to possible interactions between consecutive fluxons because of their finite width. For our intended velocity ($v \sim 0.6c$) this width is $\gtrsim \lambda_J$. 

The fluxon gates differ fundamentally from existing reversible logic
where adiabatic drive fields \cite{RenSem2011, QFPgate}
slowly evolve the gate potential while at all times retaining the state close to the potential minimum \cite{Lik1982}. 
The fluxon gates also contrast conventional SFQ gates which dissipate the potential energy 
during damped switching from the high-energy state to the low-energy state.

\subsection{Collective Coordinate Model}
\label{sec:operation_CCM}

To understand these complex nearly-ideal elastic dynamics 
we employ a collective coordinate (CC) approach \cite{DauxoisPeyrard__PhysicsOfSolitons}.
Here we briefly sketch the method and start the analysis of the 1-bit gates above. More detail on the CC approach will be provided in Sec.~\ref{sec:CCM}.

As an ansatz for the fields in left and right half of the circuit we choose
\begin{eqnarray}
\phi(x<0) &=& 
   \phi^{(\sigma, X_L)}(x) + \phi^{(-\sigma, -X_L)}(x) - 2\pi (1-\sigma) \nonumber 
   \\
\label{eq:fluxoncombination}
\phi(x>0) &=&
   \phi^{(-\sigma, X_R)}(x) + \phi^{(\sigma, -X_R)}(x) - 2\pi  
 \,,
\end{eqnarray}
each consisting of a linear superposition of fluxon and mirror antifluxon fields, where 
$\phi^{(\sigma, X)}$ is defined in \Eq{eq:soliton}. 
This conveniently parametrizes the possible asymptotic behavior of elastic scattering 
for our interest --- for one incident fluxon of polarity $\sigma$
(see insets of Fig.~\ref{fig:CCM_Epot_others}(a1)).
Asymptotically, if the coordinate $X_i$ ($i=L,R$) is far away from the interface, 
the field in the LJJ $i$, as defined in \Eq{eq:fluxoncombination},
can be either fluxon- or antifluxon-like, 
and we refer to it as a particle. 
We note that, even though the coordinate $X_L$ ($X_R$) 
can assume any value in $(-\infty,\infty)$,
the left (right) particle energy is always localized within the real space
of the left (right) LJJ, about $x_{qL} \simeq -|X_L| \leq 0$ ($x_{qR} \simeq |X_R| \geq 0$).
For $X_i = 0$, the particle excitation vanishes. 
For example, an incident fluxon in the left LJJ is represented by $X_L \ll -W$, $X_R = 0$. 
Also, a fluxon (antifluxon) in the right LJJ is given by $X_L=0$ and $X_R \gg W$ ($X_R \ll -W$).
From fits to the numerical solutions $\phi_n(t)$ we find that \Eq{eq:fluxoncombination} 
for $|X_{L,R}| \lesssim W$
also approximates the observed large interface oscillation 
(around the state $\phi(x) = 2\pi\Theta(-x)$) during gate dynamics.

By inserting ansatz \Eq{eq:fluxoncombination} in the system Lagrangian,
\Eq{eq:Lagr_orig_discrete}, 
the many degrees of freedom reduce to the two collective coordinates $X_{L,R}$. 
Using justifications above, we also constrain
$\phiBB - \phiL+\phiR = 0$ in the CC calculations, 
thereby eliminating the interface phase $\phiBB$.
This approximation is valid in a leading order perturbation expansion of the interface 
equations of motion, due to the small interface inductance
$\Lhat \ll L \lambda_J^2 / a^2$, 
as fulfilled in the simulations of Figs.~\ref{fig:4bounce} and \ref{fig:2bounce}.
We then obtain coupled equations of motion, 
\begin{eqnarray}\label{eq:EOM_CCM0}
 \left(\!\!\begin{array}{c} \ddot X_L \\[1ex] \ddot X_R \end{array} \!\!\right)
= - \mathbf{M}^{-1} \left(\!\!\begin{array}{l}
 c^2 \frac{\partial U}{\partial X_L} 
 + \frac{1}{2}\frac{\partial m_L}{\partial X_L}  \dot X_L^2 
 + \frac{\partial m_{LR}}{\partial X_R} \dot X_R^2 \\[1ex]
 c^2 \frac{\partial U}{\partial X_R}
 + \frac{1}{2}\frac{\partial m_R}{\partial X_R} \dot X_R^2 
 + \frac{\partial m_{LR}}{\partial X_L} \dot X_L^2 
 \end{array} \!\!\right)
 ,\quad
\end{eqnarray}
where the mass matrix $\mathbf{M}$ is given in Sec.~\ref{sec:CCM}. 
The diagonal mass matrix components $m_i(X_i)$ ($i=L,R$), 
describe the particle mass which varies near the interface ($X_i=0$). 
The off diagonal mass (coupling) components $m_{LR}\propto C_J^B g_I(X_L) g_I(X_R)$
approach zero away from the interface, with $g_I(X_i) = 4(\lambda_J/W) \sech(X_i/W)$.
The mass coupling is proportional to the center JJ capacitance $C_J^B$, 
and this is a key to coupling between a particle in the left LJJ and the right one.

\begin{figure}
\includegraphics[width=\columnwidth]{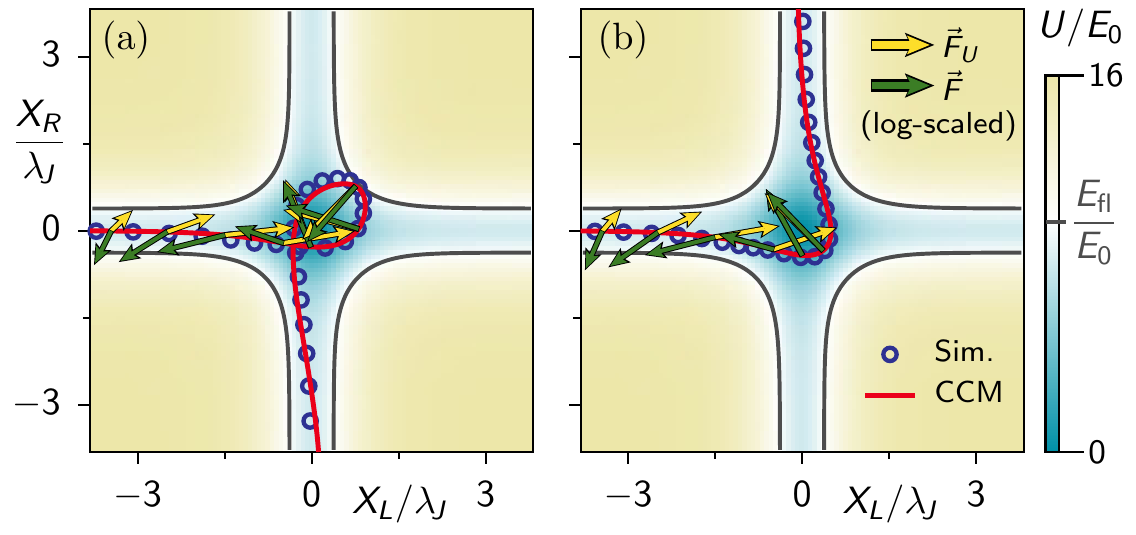}
\caption{
Trajectories $(X_L,X_R)(t)$ for interfaces of (a) Fig.~\ref{fig:4bounce} (ID)
and (b) Fig.~\ref{fig:2bounce} (NOT).
(a-b)  Identical potentials $U(X_L,X_R)$ from \Eq{eq:U_CCM0}.
Trajectories obtained from collective coordinate (CC) equations of motion, \Eq{eq:EOM_CCM0}, (red line)
and from fit of simulated $\phi_n(t)$ with \Eq{eq:fluxoncombination} (blue marker).
Arrows show accelerations (mass-normalized forces) in \Eq{eq:EOM_CCM0},
$\vec{F}_U$ from the potential (yellow arrows) 
and the total force $\vec{F} = \vec{F}_U + \vec{F}_M$ (green).
Both $\vec{F}_U$ and $\vec{F}$ differ substantially between (a) and (b) 
even initially, despite still comparable positions $(X_L,X_R)(t)$, 
e.g. first three arrows near $X_R=0$ have visibly different arrow lengths despite logarithmic scaling. 
This is caused by larger mass-matrix elements in (b), where $C_J^B$ is double of (a),  
and results in stronger initial acceleration in $-X_R$-direction. 
This causes a larger deflection of the trajectory from $X_R=0$, as well as a slowed down forward evolution (trajectory points) relative to (a).
Within the central potential well accelerations are exponentially larger and dominated by $\vec{F}_U$.
}
\label{fig:CCM}
\end{figure}

The potential $U(X_L,X_R)$ 
is symmetric under the coordinate exchange $X_L \leftrightarrow X_R$ and is of the form
\begin{equation}\label{eq:U_CCM0}
 U = U_0 + \frac{\IJab-I_c+\IJbb}{I_c\lambda_J/a} u_1
 + \frac{\IJbb}{I_c\lambda_J/a} u_{2}
 \,.
\end{equation}
Herein $U_0(X_L,X_R)$ and $u_1(X_L,X_R)$ have even parity 
under each of the transformations $X_{i} \leftrightarrow - X_{i}$ ($i=L,R$), 
while $u_{2}(X_L,X_R)$ has no parity symmetry.
One potential landscape $U$ is shown in Figs.~\ref{fig:CCM}(a) and (b).
These correspond to the ID and NOT gates of Figs.~\ref{fig:4bounce} and \ref{fig:2bounce}, 
respectively, and are identical since both gates have the same critical currents $\IJab$ and $\IJbb$. 
For either of $|X_{L,R}| \gg W$, the potential forms four valleys,
each corresponding to a single fluxon as described earlier. 
A central well exists at $(X_L,X_R) \approx (0,0)$.
Due to $\IJbb, |\IJab-I_c| \ll I_c$ the interface potentials $u_{1,2}$ contribute 
only weakly to $U$, and the valleys are therefore connected with the central well 
below the initial fluxon energy $E_{\text{fl}}/E_0 = 10$ (gray equipotential line). 
Thus, in principle conservative scattering between valleys is possible. 
The gates of Fig.~\ref{fig:CCM}(a) and (b) differ by {\it capacitance} and thus mass matrix alone, but have identical CC potential with negligible $u_2$-contribution. 
We note, however, that the parity-breaking contribution of $u_2$ can also be used to change the gate type from ID to NOT.
This is done by increasing $\IJbb$ and is discussed in Sec.~\ref{sec:CCM}.

In Figs.~\ref{fig:CCM}(a) and (b) the CC trajectories $(X_L,X_R)(t)$ 
are shown (red lines) for the ID and NOT gates, as
obtained from integration of \Eq{eq:EOM_CCM0} with initial conditions 
$X_L = X_0 \ll -W$, $\dot X_L = v_0$, and $X_R=\dot X_R =0$. 
These can be compared with trajectories obtained 
from accurate fits of the ansatz (\Eq{eq:fluxoncombination}) to the numerical simulation data 
from Figs.~\ref{fig:4bounce} and \ref{fig:2bounce} (blue markers).
From the good agreement between the two we conclude that the CC equations of motion accurately produce the correct forward-scattering dynamics.
In Fig.~\ref{fig:CCM}(a) the trajectory exhibits a net angular rotation of $3\pi/2$ into the valley with  $X_R<0$, corresponding to a fluxon emitting into the right LJJ (ID gate);
in Fig.~\ref{fig:CCM}(b) a $(\pi/2)$-rotation of the trajectory into the valley with $X_R>0$ corresponds to the emission of an antifluxon (NOT gate). 
The trajectories evolve from 
initial conditions $X_R=\dot X_R = 0$ to $X_R \lessgtr 0$,
although no significant potential coupling exists between $X_L$ and $X_R$
due to negligibly small $\IJbb$.
The coupling is instead provided by the coupling mass $m_{LR}$ in \Eq{eq:EOM_CCM0},
stemming from the relatively large capacitance $\CJbb$.

In the particle interpretation originating from the ansatz, an incident fluxon-like
particle from the left LJJ changes its character at $|X_{L}| \lesssim W$ 
and also excites a particle in the right LJJ from $|X_{R}| \lesssim W$, 
see 2nd diagram of Figs.~\ref{fig:interface}(b1) and (c1)
as well as panels (b2) and (c2).
Finally, the left particle remains at $X_{L} = 0$, while the right one
(initially at $X_{R} = 0$) exits ($|X_{R}| \to \infty$) with fluxon- or antifluxon-like character,
see last diagram of Figs.~\ref{fig:interface}(b1) and (c1).
We will analyze these dynamics in detail in Sec.~\ref{sec:CCM_gates},
after giving more detail of the CC (Sec.~\ref{sec:CCM_derivation})
and discussing general CC dynamics observed for different interface parameters
(Sec.~\ref{sec:CCM_Epot_others}).

The logic allows a comparison to billiard ball logic \cite{FredTof1982}, 
which ideally can perform reversible computing gates in two spatial dimensions (2D), 
using perfect hard collisions with barriers and each other. 
Standard scattering in a single sine-Gordon (or LJJ) system seems unlikely since there is only a delay from weak interactions between solitons (fluxons) which induces only time delays from ballistic collisions.
However, our 1-bit fluxon gates introduce elastic collisions at the interfaces due to a strongly coupled interface between the LJJ, which are confined to 1D segments (LJJs).
Rather than altering particle paths, as in the billiard ball model, they can change the particle type (polarity) during a collision. 
The collisions are  accurately described by the above particle to particle collision.
The excitations of an input particle from $|X_{L}| \ll W$ interacts with the other particle at $X_{L,R} \simeq 0$, and induces scattering to $|X_{R}| \gg W$.

\subsection{2-bit gate}
\label{sec:operation_2bit}

\begin{figure}[tb]
\includegraphics[width=\columnwidth]{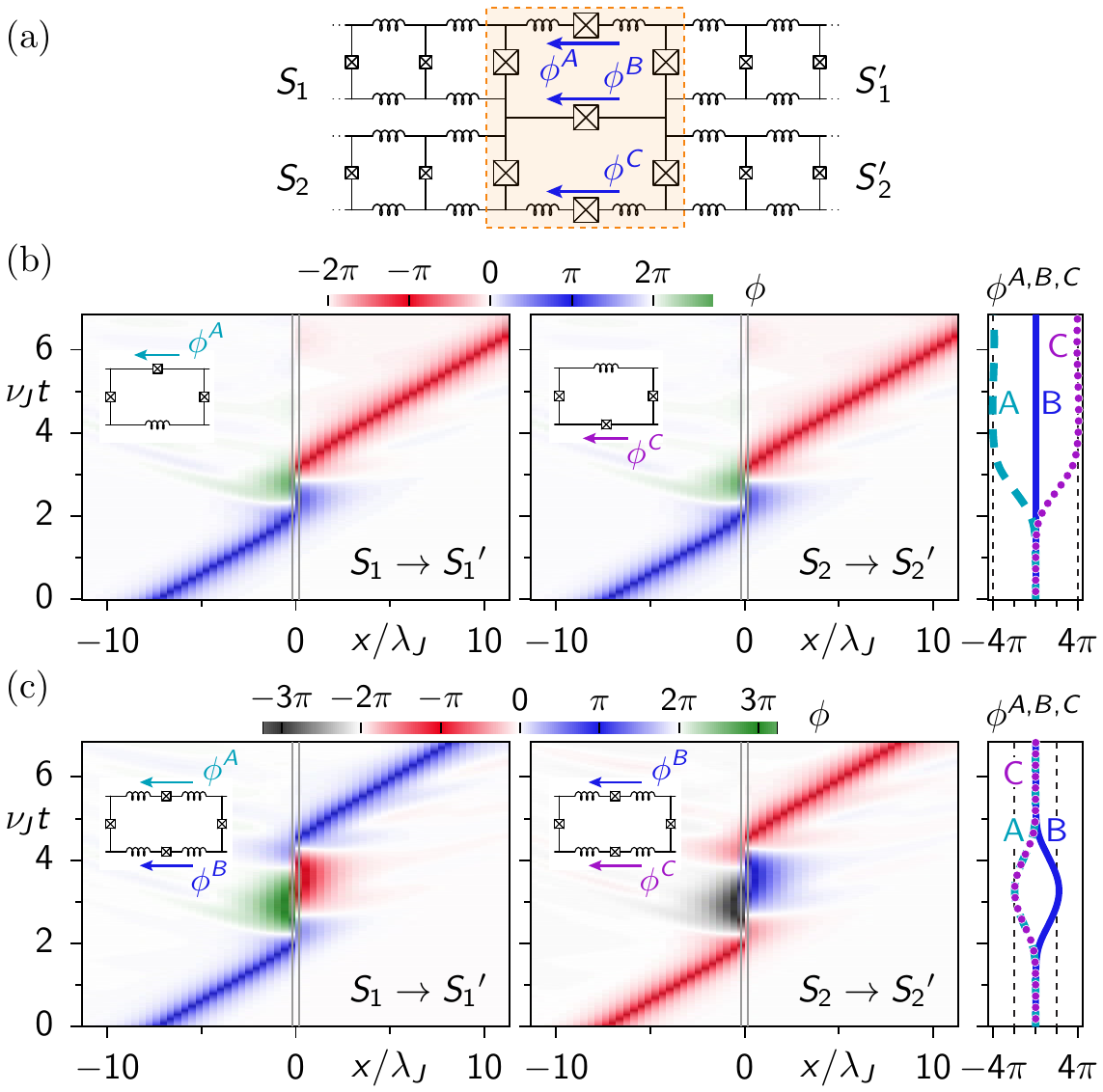}
\caption{
(a) Schematic of 2-bit NSWAP gate connecting input LJJs ($S_1$ and $S_2$) and output LJJs ($S_1'$ and $S_2'$), with left--right and vertical symmetry (the interface in the dashed box has equal A- and C-junctions).
Labeling as in Fig.~\ref{fig:4bounce}, except here two 
inductors flank upper and lower interface junctions, each with total inductance $\Lhat_A=\Lhat_C$.
(b,c) Color representation of $\phi_n$ vs.~$x_n$ and $t$ in LJJ 1 (left) and LJJ 2 (center),
and interface phases $\phi^{A,B,C}$ vs.~$t$ (right), 
for input fluxons with $v_0 = 0.6 c$. 
For input fluxons with (b) equal polarity the currents across the center interface JJs ($\phiBB$) cancel 
such that an equivalent 1-bit gate (inset and Fig.~\ref{fig:interface_AA1BB1CC1}(b)) would have 1 center JJ. 
For input fluxons with (c) opposite polarity the currents across $\phiBB$ add; an equivalent 1-bit gate exists with 2 center JJs (inset and Fig.~\ref{fig:interface_AA1BB1CC1}(c)). 
Note that center currents are predominantly through capacitors, e.g.~$C_J^{A}$ and $C_J^{B}$.
Interface parameters 
$C_J^{A}/C_J = 11.2$, $C_J^{B}/C_J = 22.0$, $\CJab / C_J = 5.82$,
$I_c^{A}/I_c = 0.02, I_c^{B}/I_c = 0.01$, $\IJab / I_c = 0.53$,
and $\Lhat^A/L = 0.30$. 
}
\label{fig:2bit_gate}
\end{figure}

We next exploit the different 1-bit gate dynamics 
for the design of a conditional 2-bit gate,
which depends on the interaction of the fields created by the input fluxons. 
This gate circuit is shown in Fig.~\ref{fig:2bit_gate}(a).
It has vertical mirror symmetry
as well as mirror symmetry along the propagation direction (left--right). 
For optimized interface parameters, as given in the caption,
near-elastic forward scattering takes place for synchronized input fluxons.
Elastic behavior here implies that in both cases, equal and opposite polarity input, 
nearly all of the energy is returned such that the output velocity nearly equals the input velocity.
The output depends on the polarities of the input fluxons:
both undergo polarity inversion if they are of the same polarity, 
as shown in Fig.~\ref{fig:2bit_gate}(b),
but input fluxons of opposite polarity both retain their original polarities,
as shown in Fig.~\ref{fig:2bit_gate}(c). 
When encoding bit states by polarities this yields a controlled NOT(SWAP) = NSWAP gate -- 
state pairs (0,0) and (1,1) are reversibly converted into each other, 
while state pairs (0,1) and (1,0) remain invariant. 
Similar to the 1-bit gates, only 2.1\% of the initial fluxon energy is dissipated in any of the NSWAP operations.
Also, with a change of the interface cell topology, through the use of wiring crossovers, the NSWAP is converted into a SWAP.
We note that, even though such 2-bit gates could also be obtained by mere rerouting (wiring crossover), 
the NSWAP gate presented here is fundamentally different since it relies on the strong interaction between the bit states at the interface. 
This is the first discovered 2-bit gate of the fluxon logic type; other 2-bit RFL gates have been subsequently investigated \cite{WusOsb2018}.

\begin{figure}[tb]
\includegraphics[width=\columnwidth]{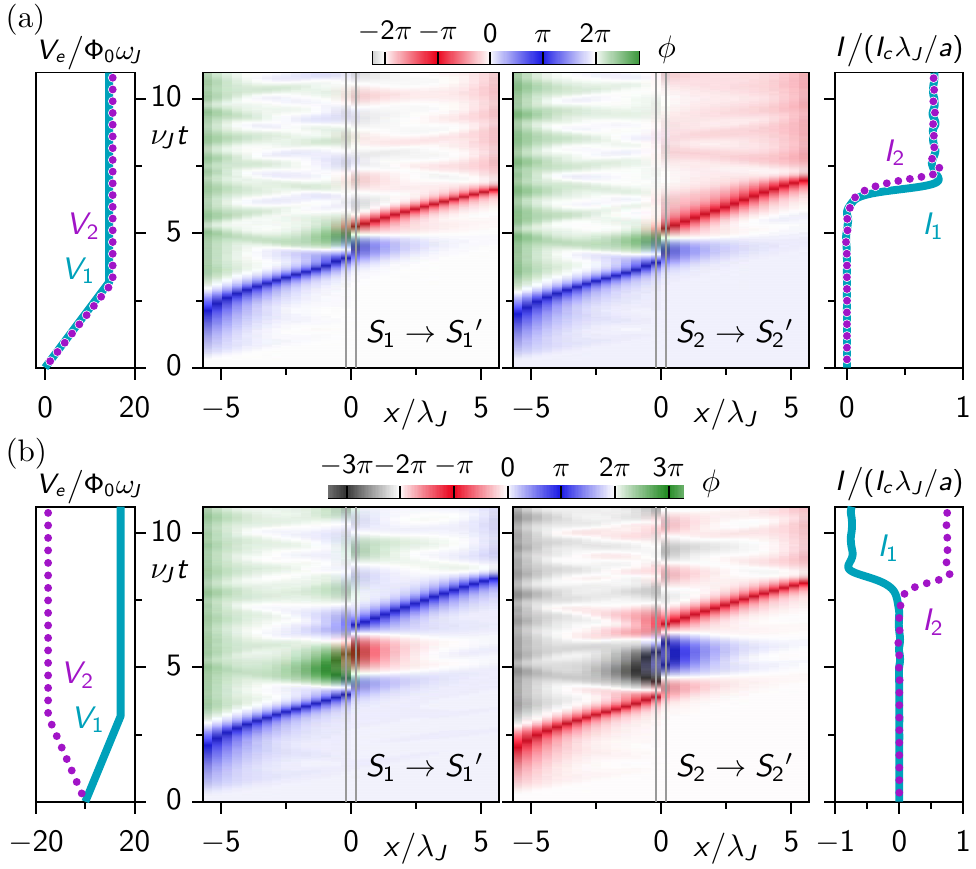}
\caption{
NSWAP gate dynamics of test platform in Fig.~\ref{fig:interface}(e):
external voltages $V_{1,2}(t)$ applied to LJJs 1 and 2 (left), 
color representation of $\phi_n$ vs.~$x_n$ and $t$ in LJJ 1 (center left) and LJJ 2 (center right), 
and currents $I_{1,2}$ in storage loops for readout (right), 
for (a) equal and (b) opposite signs of ramped external voltages. 
The interface parameters are the same as in Fig.~\ref{fig:2bit_gate}.
The added phase fluctuations relative to Fig.~\ref{fig:2bit_gate} 
is primarily related to 
small LJJ dimensions (each LJJ consists of 14 JJs),
capacitive coupling to ground (while interface not optimized under this condition),
and small imbalance between external voltages ($\max(|V_2|) = 1.06 \max(|V_1|)$).
The circuit parameters, consistent with Nb-fabrication, 
are $C_J=125\ufF, L=62.8\upH, \nu_J=21.5\uGHz, \nu_J T_{\text{ramp}}=3.2$;
the TL-LJJ coupling has $L_c=160\upH$ (both TL and LJJ loop) and 20\% relative mutual inductance;
the storage loops have $L_t = 20 L$ and $R_t=\sqrt{L/C_J}$;
each LJJ-node is coupled to ground with capacitance $C_g = 0.05C_J$.
}
\label{fig:2bit_gate_testplatform}
\end{figure}

The dependence of the interaction in the interface cell on the relative polarities of the input fields can be understood from the dynamical equivalence with 
1-bit gates.
We note that the value of $\CJab$ here is similar to that of the previously presented 1-bit gates,
which allows comparisons of the top or bottom part of the 2-bit gate to the previously discussed 1-bit dynamics.
If the input fluxons have the \emph{same polarity}
the center interface current vanishes for symmetry reasons, 
and the center junction is not excited, $\phi^{B} = 0$.
Here the dynamics of the upper and lower part of the interface are thus 
equivalent (see Fig.~\ref{fig:2bit_gate}(b) insets) to two individual 1-bit interfaces, each with a center junction 
equal to the upper center junction ($\CJaa$, $\IJaa$) of the 2-bit interface in Fig.~\ref{fig:2bit_gate}(a).
These parameters closely agree with those of the center junction of Fig.~\ref{fig:2bounce}
and thus the dynamics correspond to a NOT gate for each input fluxon. 
On the other hand, if the input fluxons have \emph{opposite polarities}
the interface dynamics in the upper and lower part are effectively decoupled in another way.
In this case, the upper and lower part of the interface are each dynamically equivalent to a 1-bit interface that has two center junctions 
rather than the one discussed previously (see insets to Fig.~\ref{fig:2bit_gate}(c), 
compare also to Fig.~\ref{fig:interface_AA1BB1}).
One of the equivalent center junctions is the same as the upper center junction of the 2-bit interface with values ($\CJaa$, $\IJaa$).
The second equivalent center junction must carry half the current of the 2-bit interface junction. 
As a consequence it must have values $(\CJbb/2$, $\IJbb/2)$ 
in order to produce the same phase winding $\phi^B$.
The critical currents $I_c^{A,B}$ are small such that we neglect them for this discussion.
We see from the the numerical values of our 2-bit gate that
$\CJbb/2 \approx \CJaa$  
which explains that $\phi^{A} \approx -\phiBB$ during the dynamics.
It is intuitive to see that this can be approximately simplified further 
to a 1-bit interface with one center junction
because the two center junctions are in series and thus act as a single interface junction with capacitance $\approx \CJaa/2 \approx 6.1 C_J$
(also described in Fig.~\ref{fig:interface_AA1BB1CC1} below).
This is in close agreement with Fig.~\ref{fig:4bounce}, where the center interface junction has the capacitance $\CJbb = 6.0 C_J$.
Therefore, as expected from analysis, the ID gate is generated for each of the input fluxons of opposite polarity. 

Continuing the comparison with billiard ball logic started above, we see that interactions between fluxons allow them to switch their polarity conditionally, while billiard balls conditionally change their paths due to collisions. Here the length scale of the interaction is defined by the Josephson penetration depth $\lambda_J$.  
Furthermore, this 2-bit gate shares features with 1-bit gates, such as large center-interface capacitance which allows the oscillatory dynamics for reversible and non-dissipative fluxon gates.

Finally we have tested the NSWAP gate in a realistic test platform, 
as shown in Fig.~\ref{fig:interface}(e). 
Figure \ref{fig:2bit_gate_testplatform} shows the results of the simulation. 
A fluxon is launched by ramping up the external voltages $V_{1,2}(t)$ of transmission lines which are inductively coupled to the input LJJs. 
The gate parameters are identical to those in Fig.~\ref{fig:2bit_gate},
however a ground plane is simulated by adding a stray capacitance to ground at each side of a junction, with a value $C_g = 0.05 C_J$.
The larger phase fluctuations relative to the simulations of Fig.~\ref{fig:2bit_gate} 
are related to small LJJ dimensions (each LJJ consists of 14 JJs),
capacitive coupling to ground (while interface not optimized under this condition),
and small imbalance between external voltages ($\max(|V_2|) = 1.06 \max(|V_1|)$).
In this test platform the output fluxons are stored in loops with large inductance $L_t$ while parallel resistances $R_t$ allow the fluxons to quickly create persistent currents in the storage loops. This test platform simulation shows the same fluxon logic despite non-ideal launch conditions (including non-equal waveforms and short left-hand side LJJs) and non-optimized gate (created by added ground plane capacitance).

As the gate examples in Sec.~\ref{sec:operation} illustrate, 
individual logic gates can operate fully ballistic, 
without external energy supply apart from the input fluxons themselves.
In Sec.~\ref{sec:optimization} the process margins of the gates will be specified 
based on a $90\%$ velocity ($95\%$ energy) retention after the gate, 
although a less stringent criterium may be practical.
Even for Nb technology this may allow a ballistic gate with 
an energy cost on the order $k_B T$, see below.
However, a logic architecture consisting of many individual gates needs to include 
clocking structures to synchronize fluxons. 
For example, in simulations of a NSWAP gate we find that the two input fluxons of equal velocity must be spaced by less than $0.4 a$ for proper gate operation.
In another work \cite{WusOsb2018} we present `store-and-launch' gates which can stop moving fluxons, store the bits as static flux states, and from these later relaunch fluxons in a synchronized way.
In the process of stopping a fluxon, it may conserve the fluxon's potential energy, which makes up $80\%$ of the total fluxon energy (for $v_0/c=0.6$).
In contrast, conventional SFQ logic gates dissipate several units of bit energy, $I_c \Phi_0$, per gate \cite{HerrETAL2011, Tolpygo2016}.
We may compare the fluxon's kinetic energy $E_K = 2 E_0$, 
which is lost when stopping the fluxon and thus forms an upper boundary 
for the energy loss in a ballistic gate, 
with the thermal energy $k_{B} T$:
Demanding that the thermal excitation of plasma waves in the LJJ 
should be suppressed,
we require that $k_B T < \hbar\omega_J$, where $\hbar \omega_J = \sqrt{E_C/E_L} E_0$ is the minimum energy of the LJJ plasma spectrum, with
$E_L = \Phi_0^2/(8 \pi^2 L)$ and $E_C = (2e)^2/(2C_J)$.
Thus we see that $E_0$ should exceed $k_B T \sqrt{E_L/E_C}$,
and we can therefore estimate that the (kinetic) energy loss of a fluxon 
in a stop-and-launch gate may be
of the order or larger than $E_K = 2 k_B T \sqrt{E_L/E_C}$.
For some realistic designs of the discrete LJJ made with Al (Nb) and for
a target temperature of $0.5\,\text{K}$ ($4\,\text{K}$), 
the factor $\sqrt{E_L/E_C}$ may be on the order $2 - 50$, 
thus resulting in a energy loss of $4 - 100 k_B T$.

\section{Collective Coordinate Model}\label{sec:CCM}

In this section we describe more details of the collective coordinate (CC) approach
leading to \Eq{eq:EOM_CCM0}, and employ it to explain elastic scattering phenomena 
in the circuit interfaces.

\subsection{Reduced (collective) equations of motion}
\label{sec:CCM_derivation}

Starting point of the analysis is the ansatz, \Eq{eq:fluxoncombination}, 
for the fields in the left and right half of the circuit 
shown in Fig.~\ref{fig:interface}(a),
each consisting of a superposition of a fluxon and mirror antifluxon.
This ansatz is consistent with the initial-state phases far away from the interface, 
namely $\phi(x \to -\infty) = 2\pi$ and $\phi(x \to \infty) = 0$ 
for a fluxon incident from the left. 
As found above, 
\Eq{eq:fluxoncombination} can be used to approximate $\phi_n(t)$ 
in the numerically simulated elastic scattering phenomena,
e.g.~for Figs.~\ref{fig:4bounce} and \ref{fig:2bounce}.
In these fits $X_{L,R}$ are independent fit parameters for each time $t$
while parameters $W$ and $\sigma$ were fixed to correspond to the initial fluxon.
According to \Eq{eq:fluxoncombination}, 
the symmetry point $(X_L,X_R)=(0,0)$ corresponds to an equilibrium field 
$\phi(x) = 2\pi \Theta(-x)$. 
Localized excitations around this state are described with $|X_{L,R}| \lesssim W$, 
some of which are sketched in the insets of Fig.~\ref{fig:CCM_Epot_others}(b1),
in particular even-parity excitations with $X_R = -X_L$
and odd-parity excitations with $X_R = X_L$ (labeled as e and o in the insets).
A full set of asymptotic scattering states are parametrized by \Eq{eq:fluxoncombination} 
as illustrated in the insets of Fig.~\ref{fig:CCM_Epot_others}(a1):
The fluxon in the left LJJ itself is represented by $X_L \ll -W$, $X_R = 0$.
Its reflection as an antifluxon is described by $X_L \gg W$, $X_R = 0$.
The forward scattering into fluxon and antifluxon 
are described with $X_R \ll -W$ and $X_R \gg W$, respectively, with $X_L = 0$.

The following paragraphs summarize results of the CC derivation given in App.~\ref{app:CCM_simple}. 
The scattering phenomena discussed here occur in a regime of small interface inductance, $\Lhat \ll L \lambda_J^2/a^2$. 
In this limit the interface equations of motion allow
the approximation $\phiBB \approx \phiL - \phiR$.
Using this approximation and the ansatz, \Eq{eq:fluxoncombination}, 
the Lagrangian, \Eq{eq:Lagr_orig_discrete}, can be written in a dimensionless form, $\tilde{\mathcal{L}} = \mathcal{L}/E_0$, as
\begin{eqnarray}
\label{eq:L_CCM}   
&& \mkern-36mu \tilde{\mathcal{L}} 
= \frac{m_L}{2} \frac{\dot{X}_L^2}{c^2} + \frac{m_R}{2} \frac{\dot{X}_R^2}{c^2} 
   + m_{LR} \frac{\dot{X}_L \dot{X}_R}{c^2} - U(X_L,X_R) 
\end{eqnarray}
with the dimensionless particle potential $U$,
\begin{eqnarray}
\label{eq:U_CCM}   
&&  U = U_0 + \frac{\IJab-I_c+\IJbb}{I_c\lambda_J/a} u_1
 + \frac{\IJbb}{I_c\lambda_J/a} u_{2}  \,.
\end{eqnarray}
The coordinate-dependent, dimensionless particle masses are
\begin{eqnarray}
&& m_i(X_i) = m_0(X_i) + \frac{\CJab-C_J + \CJbb}{C_J \lambda_J/a} (g_I(X_i))^2 
\end{eqnarray}
$(i=L,R)$, 
where $m_0(X_i)$ are the LJJ contributions to the mass, given in \Eq{eqA:m0_CCM}.
The interface factor
\begin{eqnarray}
\label{eq:gI_CCM}
&& g_I(X_i) = 4 \left(\lambda_J/W\right) \sech(X_i/W) 
\end{eqnarray}
characterizes the interface contribution to the masses, 
as well as the coupling mass,
\begin{eqnarray}
&& m_{LR}(X_L,X_R) = \frac{\CJbb}{C_J \lambda_J/a} g_I(X_L) g_I(X_R) 
\,.
\end{eqnarray}

The potential $U(X_L,X_R)$ 
has constituents $U_0(X_L,X_R)$, $u_1(X_L,X_R)$, 
and $u_{2}(X_L,X_R)$, which are defined in 
\Eqs{eqA:U0_CCM}, \eqref{eqA:u1_CCM}, and \eqref{eqA:u23_CCM}, 
and are illustrated in Fig.~\ref{fig:CCM_potentials}(a-c).
Recall that the LJJ-potential $U_0$ and the interface contribution $u_1$ 
have even parity under the transformations $X_{i} \leftrightarrow - X_{i}$,
for $i = L,R$, while the interface contribution $u_{2}$ has no parity symmetry.
All constituents of $U$ are symmetric under the left-right exchange $X_L \leftrightarrow X_R$,
as expected.

The Lagrangian, \Eq{eq:L_CCM}, generates the coupled equations of motion for $X_{L,R}$ 
\begin{eqnarray}
\label{eq:EOM_CCM}
&& \left( \ddot X_L\,, \ddot X_R \right)\big/c^2 = \vec{F}_U + \vec{F}_M \,,
\end{eqnarray}
which determine the accelerations $\ddot X_i$ by the mass-normalized force 
$\vec{F} = \vec{F}_U + \vec{F}_M$. 
The separate contributions to this force from the potential $U$ and from the mass gradients
are (cf. \Eq{eq:EOM_CCM0})
\begin{eqnarray}
\label{eq:CCM_FU}
&& \vec{F}_U = -\mathbf{M}^{-1} \left(\frac{\partial U}{\partial X_L}\,, \frac{\partial U}{\partial X_R} \right) \\
\label{eq:CCM_FM}
&& 
\vec{F}_M = -\mathbf{M}^{-1} 
\left(\!\!\begin{array}{l}
 \frac{1}{2} \frac{\partial m_L}{\partial X_L} \frac{\dot X_L^2}{c^2} 
 + \frac{\partial m_{LR}}{\partial X_R} \frac{\dot X_R^2}{c^2}   \\[1ex]
 \frac{1}{2} \frac{\partial m_R}{\partial X_R} \frac{\dot X_R^2}{c^2} 
 + \frac{\partial m_{LR}}{\partial X_L} \frac{\dot X_L^2}{c^2}
\end{array}\!\!\right) .
\end{eqnarray}
The mass matrix $\mathbf{M}$ is  composed of the dimensionless elements
$M_{ii} = m_i$ and $M_{i,j \neq i} = m_{LR}$ ($i,j=L,R$) given above.

\subsection{General elastic reflection and transmission}
\label{sec:CCM_Epot_others}

\begin{figure*}[tb]
\includegraphics[width=\textwidth]{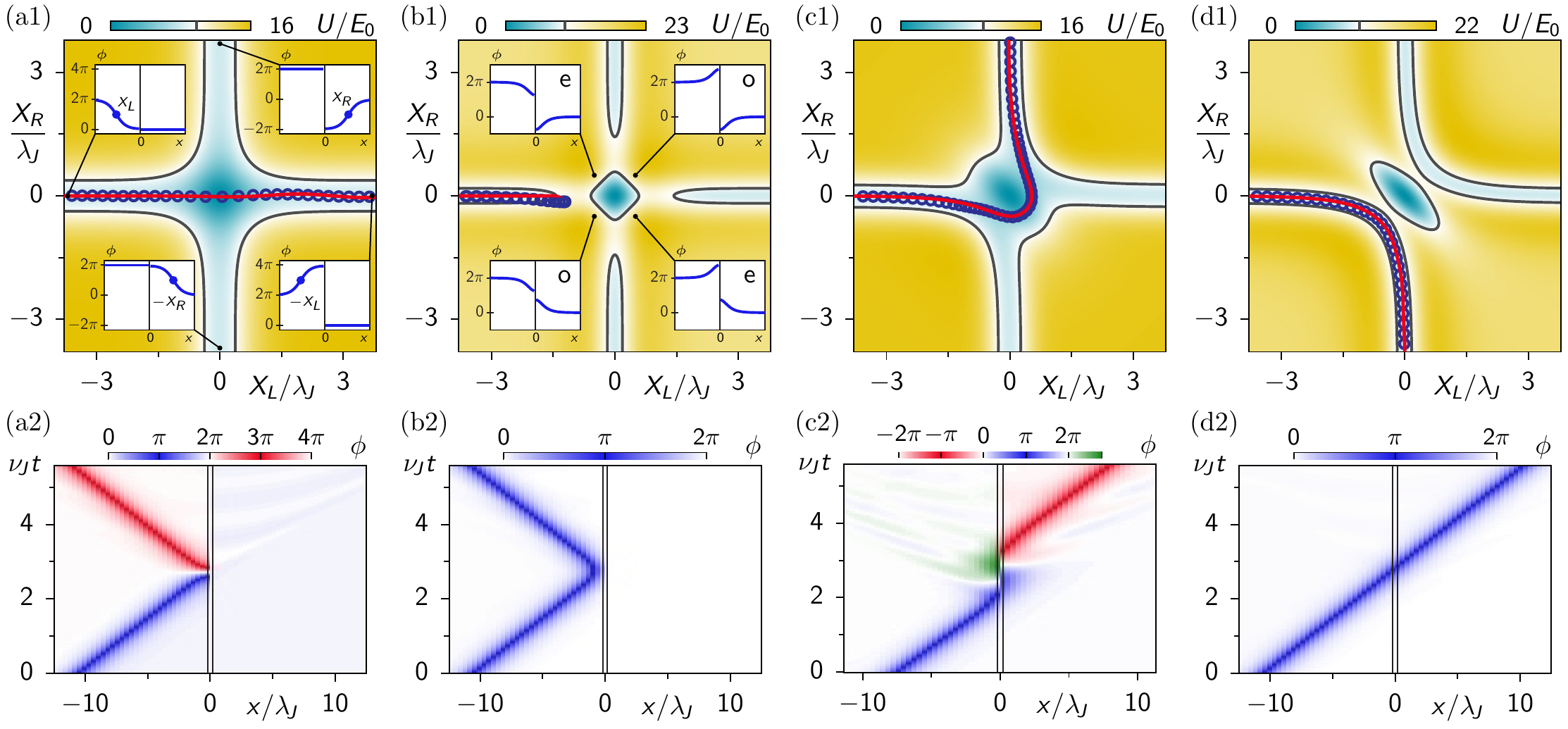}
\caption{
Elastic scattering cases in 1-bit gate structure, Fig.~\ref{fig:interface}(a), 
but with interface parameters different from gates.
Upper panels: 
Potentials $U(X_L,X_R)$, \Eq{eq:U_CCM}. Trajectories $(X_L,X_R)(t)$
from reduced (CC) equations of motion, \Eq{eq:EOM_CCM}, shown for initial condition
$\dot X_L = v_0 = 0.6 c$, $X_L \ll -W$ and $X_R=0$ (red line).
Trajectories $(X_L,X_R)(t)$ from fit of simulated $\phi_n(t)$ with \Eq{eq:fluxoncombination} (blue marker).
Lower panels: $\phi_n$ vs.~$x_n$ and $t$ from simulation for each upper panel.
Insets in (a1) and (b1) sketch CC ansatz $\phi(x)$, \Eq{eq:fluxoncombination},
evaluated at specific points $(X_L,X_R)$ and $\sigma=1$, e.g.~in (a1) clockwise from upper left: 
input fluxon, transmitted fluxon, reflected antifluxon, and transmitted antifluxon.
Interface parameters: (a,b,d) $\Lhat/L=0.10$, $\CJbb/C_J=0.10$, $\CJab/C_J=1.0$.
(a) $\IJbb/I_c = 0$, $\IJab/I_c=1.0$.
(b) $\IJbb/I_c = 0$, $\IJab/I_c=10.0$.
(c) $\IJbb/I_c = 2.9$ while all other parameters are the same as in Fig.~\ref{fig:4bounce}.
(d) $\IJbb/I_c = 10.0$, $\IJab/I_c=1.0$.
}
\label{fig:CCM_Epot_others} 
\end{figure*}

The composite potential $U(X_L,X_R)$ is illustrated in Fig.~\ref{fig:CCM_Epot_others}(a1) 
for $\IJab = I_c$ and $\IJbb = 0$, such that both $u_1$ and $u_{2}$ do not contribute.
The potential $U=U_0$ forms four scattering valleys for the asymptotically 
free fluxons either left or right of the interface (see insets),
which are connected by a central potential well at $X_L=X_R=0$.
The accessible coordinate space is limited by 
the initial fluxon energy $E_{\text{fl}}/E_0 = 10$ (gray equipotential line).
Fig.~\ref{fig:CCM_Epot_others}(a1) also shows
the trajectory $(X_L, X_R)(t)$ obtained from integration of \Eq{eq:EOM_CCM} with initial condition 
$X_L = X_0 \ll -W$, $\dot X_L = v_0$, and $X_R=\dot X_R =0$, corresponding to the incident fluxon (red line).
It is in excellent agreement with the corresponding trajectory obtained from fitting
$\phi_n(t)$ 
with \Eq{eq:fluxoncombination} (blue markers).

Here the interface capacitances are $\CJbb \ll C_J$ and $\CJab = C_J$,
such that the coupling $m_{LR}$ is negligible and the acceleration in \Eq{eq:EOM_CCM}
remains small in $X_R$-direction. 
The resulting motion from $X_L<0$ to $X_L>0$ corresponds to the fluxon being reflected
as antifluxon, as shown in Fig.~\ref{fig:CCM_Epot_others}(a2).
Note that, in accordance with an earlier remark, the left particle is throughout this process found at a real-space position $x_{qL} \simeq -|X_L| < 0$ within the left LJJ.
This scattering is more general than an open LJJ boundary which also generates antifluxon reflection, as a result of the Neumann boundary condition \cite{CostabileETAL1978} $\phi'(x=x_{\text{boundary}})=0$. Unlike that open boundary, here non-negligible current flows through the left interface junction.
We note that in this parameter regime a current-phase relation is observed for elastic reflection at the interface which relates to a general integrable boundary condition for the SGE \cite{GhoZam1994}, but which is outside the scope of the current work.

Keeping $\IJbb=0$ but raising $\IJab \gg I_c$ relative to the previous case
results in a finite prefactor to the interface potential $u_1$ and 
creates a potential barrier at $|X_i| \approx W$ around the center well as shown in Fig.~\ref{fig:CCM_Epot_others}(b1)
(compare also to $u_1$ in Fig.~\ref{fig:CCM_potentials}(b)).
In particular, if $(\IJab - I_c + \IJbb) u_1/I_c$ is larger than $E_{\text{fl}}/E_0$,
scattering between the valleys is prevented and 
the incident particle 
is reflected before entering the central potential well,
as shown in Fig.~\ref{fig:CCM_Epot_others}(b2).
This is similar to the simple case of a fluxon reflection at a shunted end of a LJJ 
(with the Dirichlet boundary condition \cite{VanDuzer} 
$\phi(x=0) = 0$),
in that the particle will remain fluxon-like. 

We next consider $\IJbb>0$ where the contribution of the interface potential $u_{2}$ 
breaks the even potential parities, 
as shown in Fig.~\ref{fig:CCM_Epot_others}(c1).
The trajectory coming in at $X_R=0$ is therefore subject to a relatively strong acceleration component $F_{U,R}$ by which at first it is deflected towards $X_R<0$, before moving into the valley with $X_R>0$.
Note that here we have additionally set large values of $\CJbb, \CJab$, 
which modify $\vec{F}_U$, and also provide a strong mass-gradient acceleration $\vec{F}_M$.
Under their combined action the trajectory eventually evolves smoothly into the valley at $X_R > 0$, corresponding to an antifluxon released into the right LJJ, 
as shown in Fig.~\ref{fig:CCM_Epot_others}(c2).
This forms an alternative NOT gate to Fig.~\ref{fig:2bounce} (Fig.~\ref{fig:CCM}(b)).
In this alternative, only $\IJbb$ differs from the interface parameters of the ID gate in Fig.~\ref{fig:4bounce} (Fig.~\ref{fig:CCM}(a)), 
and therefore the altered potential is responsible 
for turning the ID into a NOT gate.
Recall the gates of Fig.~\ref{fig:CCM} differ only by $\CJbb$, 
with negligibly small $\IJbb$, and thus does not make use of this parity-breaking effect.

If $\IJbb$ increases further the odd parity contribution in $u_{2}$
disconnects the center potential well from the valleys, 
as shown in Fig.~\ref{fig:CCM_Epot_others}(d1).
However, the asymptotic valleys of equal-polarity fluxons become simply connected: $(X_L \ll 0,X_R=0)$ with $(X_L=0,X_R \ll 0)$ and $(X_L \gg 0,X_R=0)$ with $(X_L=0,X_R \gg 0)$. 
The trajectory is confined to a curved valley which corresponds to the direct transmission of the fluxon across the interface, 
as shown in Fig.~\ref{fig:CCM_Epot_others}(d2).

\subsection{Two NOT gate types, made from one ID gate}
\label{sec:CCM_gates}

Finally, in Figs.~\ref{fig:CCM}(a) and (b) the potential $U(X_L,X_R)$ and 
trajectories $\left(X_L,X_R\right)(t)$ are illustrated 
for the interface parameters underlying the 
gates of Figs.~\ref{fig:4bounce} and~\ref{fig:2bounce}.
Recall that both cases have the same potential $U$
because the interface parameters $\IJbb, \IJab$ are identical. 
The trajectories demonstrate how the unconventional fluxon dynamics of these cases can be attributed to competition between potential and mass-gradient accelerations, 
$\vec{F}_U$ (yellow arrows) and $\vec{F}_M$,
which create the total acceleration, 
$\vec{F} = \vec{F}_U + \vec{F}_M$ (green arrows).
Forward scattering of the fluxon requires a deflection of the trajectory, initially at $X_R=0$,
into one of the valleys at $X_R \neq 0$.
The gates are based on interfaces with very small $\IJbb/I_c$, 
such that the potential $U$ has approximately even parities. Therefore, 
a non-negligible mass coupling $m_{LR} \propto \CJbb$ in \Eq{eq:EOM_CCM} 
is required to deviate from $X_R=0$. 
First, while approaching the central potential well, $X_L \ll -W$, 
the trajectory is deflected from $X_R = 0$ by the dominant action of 
$\vec{F}_M$, as shown by the green $\vec{F}$-arrows which point in a different direction than the yellow $\vec{F}_U$-arrows.
In this limit, and for initially $X_R=\dot X_R=0$,
we calculate the $X_R$-component of $\vec{F}_M$ as 
\begin{equation}\label{eq:FMR_init}
 F_{M,R} 
%
\approx 
 -\frac{\dot X_L^2}{\omega_J^2 W^2}\,
 \frac{2\frac{\CJbb}{C_J \lambda_J/a}}{1 + \frac{\lambda_J}{W} \frac{\CJab - C_J + \CJbb}{C_J \lambda_J/a}} \,  e^{X_L/W} 
\,.
\end{equation}
Though exponentially suppressed for $X_L \ll -W$,
$F_{M,R}$ initially dominates compared to the other acceleration components, 
which scale as
$|F_{U,R}| \propto e^{3 X_L/W}$ and 
$|F_{U,L}|, |F_{M,L}| \propto e^{2 X_L/W}$.
$F_{M,R}$ deflects the incoming trajectory towards $X_R<0$. 
Then, once it enters the central potential well at $|X_L| < W$
this small deviation $X_R<0$ experiences a strong restoring force dominated by $\vec{F}_U$.

In Fig.~\ref{fig:CCM}(b) (the NOT gate) the initial deflection is larger than
in Fig.~\ref{fig:CCM}(a) (the ID gate) due to the larger value of $\CJbb$. 
Note that the force arrows in Fig.~\ref{fig:CCM} are scaled logarithmically, 
such that the initial green acceleration arrow in (b) appears only slightly longer compared to (a), 
while $F_{M,R}$ in fact is a factor $1.4$ larger in (b) according to \Eq{eq:FMR_init}. 
The result of larger initial acceleration counteracting the motion in (b)
is seen in the lower speed of the particle at $X_L<-W$, 
as indicated by the smaller separation of the adjacent markers 
representing the trajectories at equal time steps. 
The force balance is designed here to symmetrically reflect the trajectory 
in the potential well at a point on the symmetry line $X_R = -X_L < 0$
(the velocity-component along that direction vanishes). 
The result is a total $(\pi/2)$-counter clockwise rotation into the valley with $X_R > 0$,
as described above.

In Fig.~\ref{fig:CCM}(a) the initial deflection of the ID gate 
is reduced by a factor $1.4$
due to the half value of $\CJbb$ (see \Eq{eq:FMR_init}) relative to 
Fig.~\ref{fig:CCM}(b). 
The trajectory thus retains larger velocity in $X_L$-direction when entering the central potential. It is then reflected symmetrically across the symmetry line $X_R=X_L>0$, related to a $(3\pi/2)$-rotation. 
In this case $E_{\text{fl}}/E_0 \gtrsim U(X_L(t), X_R(t))$ near the symmetry line  and thus the velocity-component perpendicular to this line is also small. 
This results in slower evolution of the phases compared with the NOT gate, 
and together with the longer trajectory in the central potential well implies a longer gate time of the ID gate.

In contrast to Fig.~\ref{fig:CCM}(a) and (b), 
where a conversion from ID to NOT gate is achieved by  
a change of the interface parameter $\CJbb$ alone, 
this conversion can also be achieved with a different parameter change ---
by tuning the interface potentials $u_1$ and $u_2$ through $\IJbb$.
This was discussed in Sec.~\ref{sec:CCM_Epot_others}, 
where the increase from $\IJbb/I_c \ll 1$ to $\IJbb/I_c \approx 3$
generates the NOT gate of Fig.~\ref{fig:CCM_Epot_others}(c2).
The corresponding CC potential and trajectory are shown in Fig.~\ref{fig:CCM_Epot_others}(c1).
In this example, the initial deflection to $X_R<0$ is enhanced compared to the ID gate 
(Fig.~\ref{fig:CCM}(a)) by the parity-breaking influence of the interface potential $u_2$
(other than $\IJbb$ the parameters are the same).
Furthermore, the trajectory is slowed down at $X_{L} \approx -W$ due to a potential barrier 
introduced by $u_1$, such that the trajectory is symmetrically reflected already 
across the symmetry line $X_R=-X_L<0$.

\section{Gate optimization and robustness}
\label{sec:optimization}


%
\begin{figure}[tb]
\centering
\includegraphics[width=1.0\columnwidth]{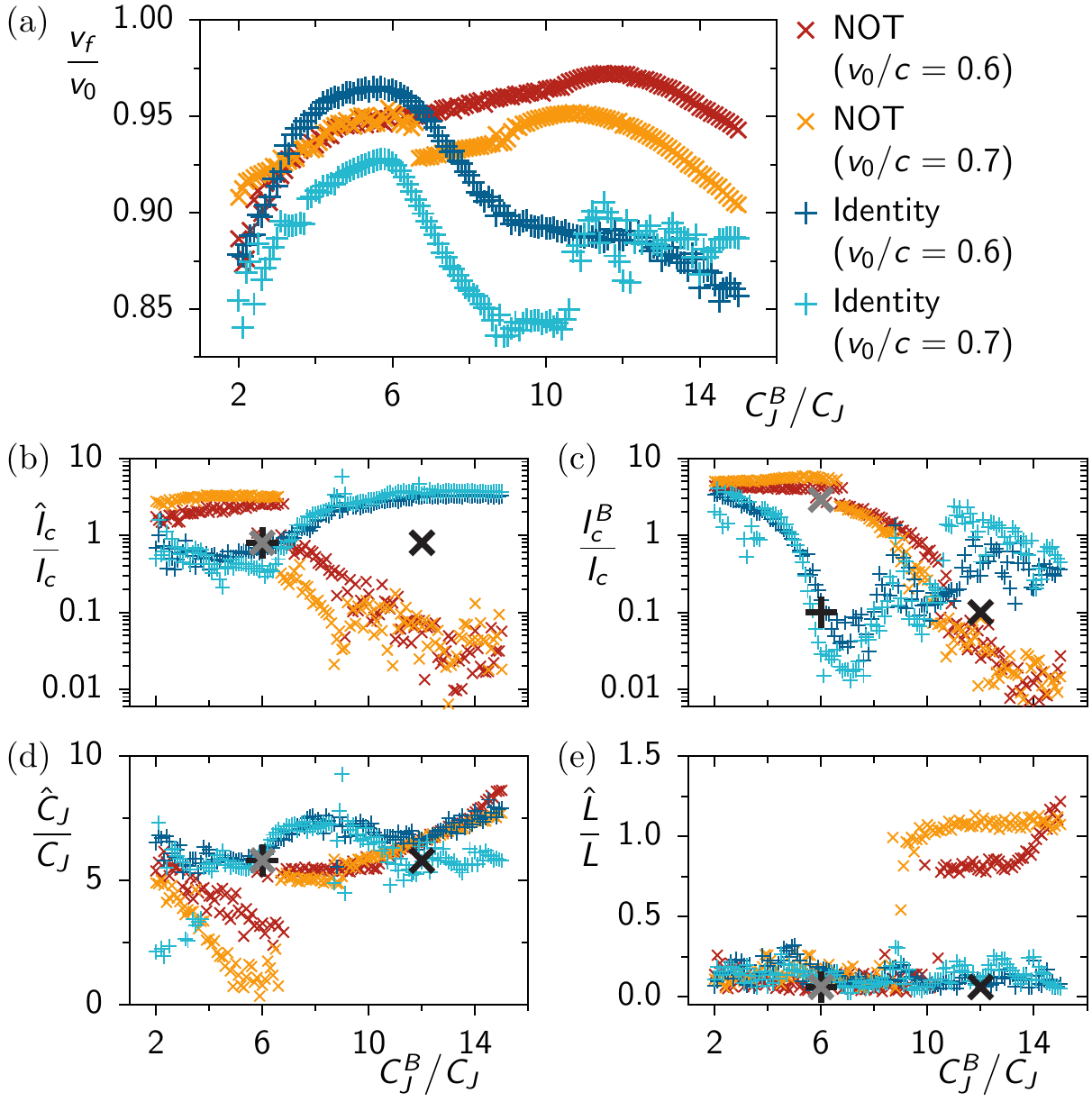}
\caption{
Interface parameters supporting near-elastic forward scattering for 1-bit gates.
(a) Final-to-initial velocity ratio $v_f/v_0$,
for interface with $\CJbb$ and with (b-e)
values of remaining interface parameters [$\IJab, \IJbb, \CJab, \Lhat$], 
obtained from numerical optimization of $v_f/v_0$.
Optimized parameter sets are shown for NOT gate, with initial velocities
$v_0/c = 0.6$ (red {\sf x}) and $v_0/c = 0.7$ (orange {\sf x}),
and for ID gate, with $v_0/c = 0.6$ (blue {\sf +})
and $v_0/c = 0.7$ (light blue {\sf +}).
Additional large markers in (b-e) indicate gate parameters of: 
ID from Fig.~\ref{fig:4bounce} (black {\sf +}), 
NOT from Fig.~\ref{fig:2bounce} (black {\sf x}), 
and NOT from Fig.~\ref{fig:CCM_Epot_others}(c) (grey {\sf x}).
Not all parameters need to be close to their optimum for acceptable gate performance,
such as $\IJab$ in the NOT gate of Fig.~\ref{fig:2bounce}.
}
\label{fig:optimparams__CJbb}
\end{figure}

The CC analysis based on \Eq{eq:fluxoncombination} was used to explain 
the role of interface parameters $\CJbb$, $\IJbb$, and $\IJab$ in the dynamics. 
For example, in the 1-bit gates 
a moderately large capacitance $\CJbb$ is required for the mass coupling $m_{LR} \propto \CJbb$
to generate a non-negligible initial acceleration
in the $X_R$-direction $F_{M,R}$, as given by \Eq{eq:FMR_init}.
To achieve gate efficiency beyond the CC predictions
Monte Carlo optimizations were performed, 
based on the full simulation with \Eq{eq:Lagr_orig_discrete},
where we vary $\CJbb > C_J$ as a independent variable. 
For each value of $\CJbb$ we maximize the final velocity $v_f$ 
of the fluxon emitted into the right LJJ
through random iterations of $\IJab, \IJbb, \CJab$, and $\Lhat$.
Fig.~\ref{fig:optimparams__CJbb}(a) shows the resulting optimized $v_f$,
obtained with interface parameters presented in Figs.~\ref{fig:optimparams__CJbb}(b-e).
This is shown for initial velocities $v_0/c = 0.6$ and $v_0/c = 0.7$, 
both for the ID and NOT gates.

For the NOT gate a broad maximum of $v_f/v_0 \approx 97\%$ is found 
for $v_0/c = 0.6$ at $\CJbb/C_J \approx 12$,
and of $v_f/v_0 \approx 95\%$ for $v_0/c = 0.7$ at $\CJbb/C_J \approx 11$.
At even smaller velocities the maximum is shifted further to higher $\CJbb/C_J > 12$.

The optimized ID gate 
has a narrower peak in $v_f$ in the displayed range than the NOT gate, 
with a maximum at $\CJbb/C_J \approx 6$, 
reaching $v_f/v_0 \approx 97\%$ for $v_0/c=0.6$ and $\approx 92\%$ for $v_0/c=0.7$.
The optimum $\CJbb$ is less dependent on $v_0$ for this gate. 
Around the peaks in $v_f$, e.g. $\CJbb/C_J \approx 12$ for the NOT gate,
the interface parameters are relatively constant, or else are negligibly small:

(i) 
As shown in panel (d),
for both gates the optimal $\CJab$ is larger than $C_J$ (as well as $\CJbb$). 
It has the approximate value of $\CJab \approx \CJbb/2$ for the NOT gate 
(for $\CJbb/C_J > 10$)
and $\CJab \approx \CJbb$ for the ID gate (for $5 < \CJbb/C_J < 8$).
In contrast, if $\CJab$ is strongly reduced from these optimal values
the resulting larger initial deflection towards $X_R < 0$ 
can destroy the gate dynamics, as understood from the CC analysis (cf.~\Eq{eq:FMR_init}). 
Usually, repeated large-amplitude bounces within the central potential well are then observed.
Similar to what happens when $\CJbb$ is increased (as shown in Fig.~\ref{fig:CCM}),
when $\CJab$ is reduced from the optimized ID gate of Fig.~\ref{fig:CCM}(a) 
with $\CJab/C_J=5.8$ to $\CJab/C_J = 1$
the initial deflection increases, and as a result the trajectory creates a NOT gate.

(ii) 
As shown in panels (b) and (c),
near the optimization maxima of $v_f/v_0$, 
both $\IJbb$ and $\IJab$ are small relative to $I_c$,
showing that the short-range interface potentials $u_1$, $u_2$ are 
irrelevant for the operations of the optimized gates. 
Instead these are determined by the balance of mass matrix elements, 
weighted by $\CJbb$ and $\CJab$, in the fixed potential $U \approx U_0$.
Only away from their respective optima  
do the gates increasingly appear to rely on 
the modification of the potential $U(X_L, X_R)$ by contribution of $u_1$ and $u_2$.
This is particularly evident in case of the NOT gate, where $\IJbb$ and $\IJab$
at $\CJbb/C_J < 6.5$ form an elevated plateau, with $\IJbb, \IJab > I_c$.
For $v_0/c=0.7$ these parameters even yield a sudden 
increase of $v_f$ by $\approx 2\%$ to match the performance of $v_0/c=0.6$, see panel (a).
An example for the dynamics in this regime of large $\IJbb$ and $\IJab$ 
(interface parameters indicated by large gray marker in Fig.~\ref{fig:optimparams__CJbb}(b-e)) 
is shown in Fig.~\ref{fig:CCM_Epot_others}(c).
Similarly, for the ID gate away from its optimal $\CJbb$, 
$\IJab$ is enhanced, particularly when $\CJbb/C_J > 8$. 
This allows $u_1$ to establish a narrower valley in the $X_R$-direction at $X_L \approx -W$ 
(see Fig.~\ref{fig:CCM_potentials}(b1)),
and this compensates for the larger initial $F_{M,R}$ due to $\CJbb$ in \Eq{eq:FMR_init}. 
For $\CJbb/C_J < 4$ (below the optimum) much larger values of $\IJbb$ are observed,
and this breaks the even potential parities due to the increased contribution of $u_2$.
This compensates for the smaller initial $F_{M,R}$ (arising due to smaller $\CJbb$).

(iii)
The Monte Carlo optimization generally yields $\Lhat/L \approx 0.1$, as shown in panel (e). 
This regime is consistent with the assumptions of the CC analysis, 
which uses $\Lhat/L \ll \lambda_J^2/a^2$.
A seemingly large $\Lhat \approx L$ is found in plateaus of the NOT gates at $\CJbb/C_J > 10$, 
but this still corresponds to a small expansion parameter $(\Lhat a^2)/(L \lambda_J^2) \approx 1/7$,
valid in earlier approximations. 

The similar parameters for the two gates in Fig.~\ref{fig:optimparams__CJbb} 
indicate the possibility 
to convert between the two gate types by adjusting only a single parameter.
As examples, we have discussed in Sec.~\ref{sec:CCM} 
how the change of either $\CJbb$ from $\CJbb/C_J = 6$ to $\CJbb/C_J = 12$, 
or of $\IJbb$ from $\IJbb/I_c = 0.1$ to $\IJbb/I_c = 2.9$,
can turn a near-optimized ID into a near-optimized NOT gate.
The parameters of these three interfaces are indicated in Fig.~\ref{fig:optimparams__CJbb}
with the large black and gray markers.
As already mentioned, a similar effect can be achieved by decreasing $\CJab$.

%
\begin{figure}[tb]
\centering
\includegraphics[width=1.0\columnwidth]{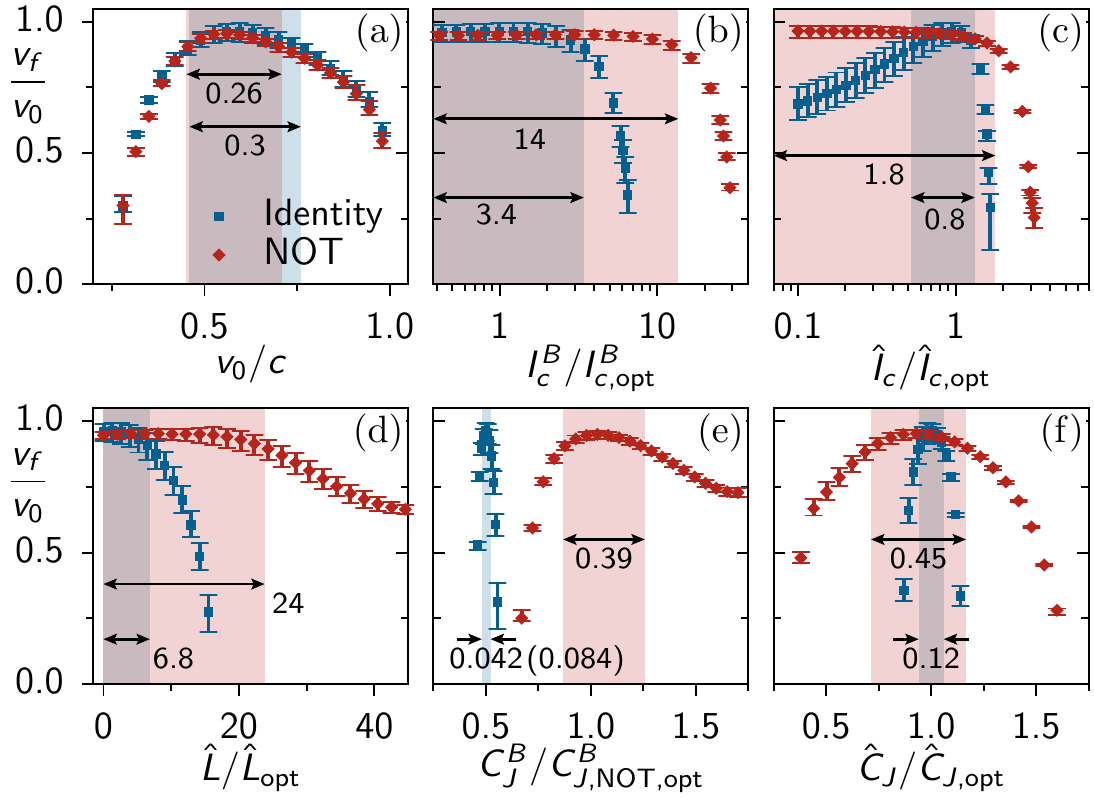}
\caption{
Robustness of 1-bit gates:
ratio of gate output to input velocity, $v_f/v_0$,
as function of (a) initial velocity $v_0$.
(b-f) ratio $v_f/v_0$ for $v_0/c=0.6$, shown as function of varied interface parameter,
which is (b) $\IJbb$, (c) $\IJab$, (d) $\Lhat$, (e) $\CJbb$ and (f) $\CJab$.
In (a) all interface parameters, 
and in (b-f) all but the varied interface parameter
are kept constant at values
$\hat{I}_{c,\text{opt}}$, $I_{c,\text{opt}}^B$, $\Lhat_{\text{opt}}$,
$C_{J,\text{opt}}^B$, $\hat{C}_{J,\text{opt}}$,  
used  in Fig.~\ref{fig:4bounce} for the ID (blue square) 
and in Fig.~\ref{fig:2bounce} for the NOT gate (red diamond), respectively. 
Error bars mark the amplitudes of velocity oscillations after scattering.
Shaded regions illustrate the ranges wherein $v_f/v_0 > 0.9$. This allows for fast output and a sufficient range of parameters for fabrication. However only $v_f/v_0 > 0$ is necessary for deterministic (error-free) gate results with potential energy conservation, a minimum requirement for incorporation into a reversible architecture (see text).
}
\label{fig:sensitivity}
\end{figure}

In Fig.~\ref{fig:sensitivity} we illustrate the gate robustness with respect to
variations in the initial conditions and the interface parameters,
for the gates of Figs.~\ref{fig:4bounce} and \ref{fig:2bounce}.
Each panel shows the final velocity of the output fluxons under one of these parameter variations.
As demonstrated in Fig.~\ref{fig:sensitivity}(a) the final-to-initial velocity ratio $v_f/v_0$ 
as a function of initial velocity $v_0$ is similar for both 
ID and NOT gates, yielding $v_f/v_0 > 90\%$ in the wide interval
$0.45 \leq v_0/c \lesssim 0.7$. 
For $v_0/c = 0.6$ this corresponds to an energy loss $\Delta E/E_{\text{fl}} < 5\%$,
while optimal forward-scattering at the maxima have $\Delta E/E_{\text{fl}} \lesssim 3\%$.
The relative insensitivity to the incident velocity is remarkable. 
Resonant scattering at defects within a LJJ, e.g.~local inhomogeneities of $I_c$ or $C_J$, 
typically exhibit high sensitivity to velocity variations, as characteristic for chaotic scattering processes \cite{FeiKivVaz1992, GooHab2007}. 
We note that our interface cannot directly be compared with a mere LJJ defect: while the latter allows a perturbation treatment within the Sine-Gordon theory, our interface involves an essential additional degree of freedom, $\phiBB$, which allows non-perturbative effects.

The robustness against relative variations 
$p/p_{\text{opt}}$ of one parameter, 
$p \in [\IJbb, \IJab, \Lhat, \CJbb, \CJab]$, 
with respect to its optimum value $p_{\text{opt}}$
is shown in Figs.~\ref{fig:sensitivity}(b-f).
The acceptable range of parameter variations seen in these figures
is related to the scatter of optimized parameters in Fig.~\ref{fig:optimparams__CJbb}, when the gate parameter (large black marker in Fig.~\ref{fig:optimparams__CJbb}(b-e)), 
lies within the scatter range. The value is in the optimized range within the scatter for the values of $\CJab$ of both gate types, Fig.~\ref{fig:optimparams__CJbb}(d).
In this case, greater scatter in the parameter is seen in the NOT gate relative to the ID gate, 
and thus as expected the NOT gate is less sensitive in the $\CJab$-gate parameter, 
as seen in Fig.~\ref{fig:sensitivity}(f).
In other cases the chosen gate parameters lie outside that scatter range because it is not very sensitive to the parameter, such as $\IJab$ in the NOT gate of Fig.~\ref{fig:2bounce}.
The NOT gate is rather insensitive to most parameter variations, 
including $\IJbb$ and $\IJab$. In line with previous analysis, $\Lhat$ is small and therefore it is relatively insensitive for both gate types.
For the ID gate the criterion $v_f/v_0 \geq 0.9$ 
gives a large accessible range of $+ 30$ to $- 48 \%$ for parameter $\IJab$.
It also gives tolerance ranges for $\CJbb$ and $\CJab$ of $8 \%$ and $12 \%$, respectively.
Note that the variation of $\CJbb$ is presented in panel (e) 
relative to the optimized value of the NOT gate, 
which is twice that of the ID gate: 
$C_{J,\text{NOT},\text{opt}}^B/C_{J,\text{ID},\text{opt}}^B = 2.0$. 
From this analysis, the narrowest tolerance range, $8\%$ in capacitance is for the ID gate, but still shows compatibility with current fabrication processes. 
Further optimization can be performed, and furthermore the tolerance range (margins) 
can increase by large factors when setting a lower output-velocity criterion.

\section{2-bit gate}\label{sec:2bitgates}

\begin{figure}[bt]
\centering
\includegraphics[width=1.0\columnwidth]{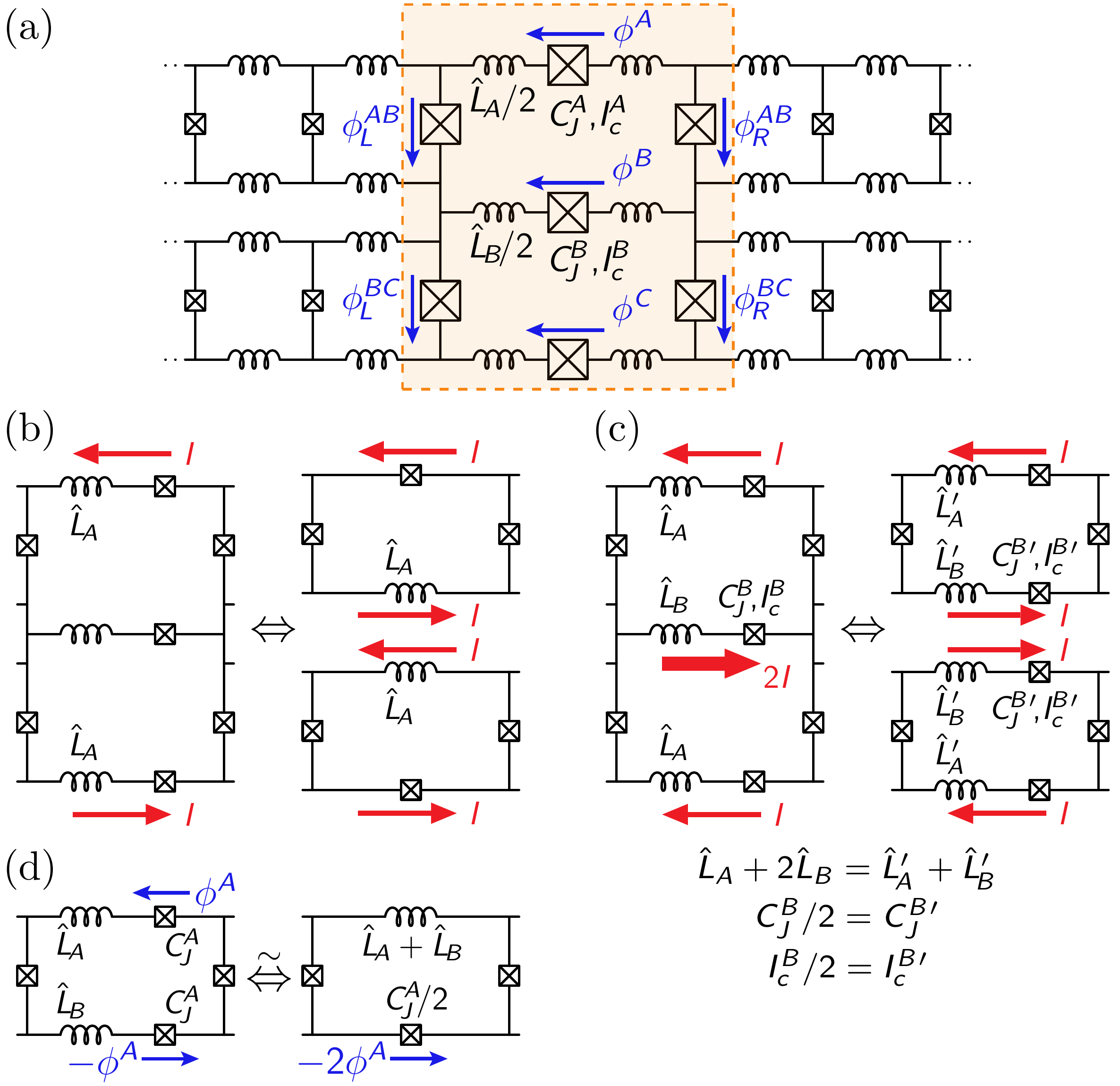}
\caption{
(a)
2-bit gate with highlighted
interface connecting two LJJs from each side. 
Vertical symmetry 
($\CJcc = \CJaa$, $\IJcc=\IJaa$, $\Lhat_C=\Lhat_A$
$\CJab^{BC} = \CJab^{AB}$, 
$\IJab^{BC} = \IJab^{AB}$)
as well as left--right symmetry.
Left--right visual symmetry (as in Fig.~\ref{fig:interface}(a)) 
is shown for $\Lhat_{A,B,C}$ but later panels combine the two 
half sized inductors for an equivalent circuit.
Dynamically equivalent 1-bit interfaces 
for input of 2 synchronized fluxons in the 2-bit circuit with: 
(b) equal or (c) opposite polarity.
(d) Approximate equivalence between 1-bit interfaces, valid for small critical currents 
$\IJaa, \IJbb \ll I_c$.
Labels for invariant parameters in mappings (b-d) are omitted. 
}
\label{fig:interface_AA1BB1CC1}
\end{figure}

Now we discuss the interface for the 2-bit gate, Fig.~\ref{fig:interface_AA1BB1CC1}(a), 
which connects two LJJs from each side. 
It is designed as a generalization of the 1-bit gate interface,
in that it is symmetric in propagation direction (left--right),
with central junctions in the interface, labeled A,B,C.
We also restrict the circuit to the case of vertical symmetry about the B-line, 
e.g.~$\CJcc=\CJaa$.
We introduce the notation $\phiAB$ and $\phiBC$ for the phases 
across the junctions in the upper and lower LJJs, respectively.

We discuss here the cases of aligned and anti-aligned input fields: 
$\phiBC_n = \phiAB_n$ and  $\phiBC_n = -\phiAB_n$, respectively. 
For synchronized input fluxons of identical velocity
the former corresponds to both having equal polarities,
and the latter to opposite polarities.
A coupling between fields of upper and lower LJJ can only occur within the interface.
The dynamics in the circuit of Fig.~\ref{fig:interface_AA1BB1CC1}
cannot be mapped (impedance matched) to dynamics in 1-bit interfaces for arbitrary input fields
due to the nonlinear circuit elements.
However, as detailed in App.~\ref{app:2bitinterface}, 
it is possible for the above special initial conditions where 
the dynamics becomes fully equivalent to the independent evolution in 1-bit circuits,
such that the initial property $\phiBC_n = \pm \phiAB_n$ remains preserved.
For simplicity we argue in the limit of small interface inductances $\Lhat_A$, $\Lhat_C$.
Similar to the first-order approximation in the CC analysis of Sec.~\ref{sec:CCM_derivation}, 
this allows us to employ the interface-cell constraints 
$\phiAA-\phiBB = \phiAB_R - \phiAB_L$ and $\phiBB-\phiCC = \phiBC_R - \phiBC_L$.
Next we schematically summarize the formal analysis provided 
in App.~\ref{app:2bitinterface} for the two initial cases.

\paragraph*{Case I, equal polarities:}
Here $\phiBC = \phiAB$ and $\dot{\phi}^{BC} = \dot{\phi}^{AB}$.
The above cell constraints then imply that $\phiAA-2\phiBB+\phiCC = 0$.
This input symmetry together with the device symmetry 
imply $\phiCC = -\phiAA$, and thus $\phiBB = 0$.
The current in the junction with $\phiBB$ thus vanishes,
while the currents in the junctions with $\phiAA$ and $\phiCC$ are opposite.
Under these conditions, the interface Lagrangian
becomes a sum of two independent contributions,
and it is therefore clear that $\phiAB$ and $\phiBC$ evolve independently.
Because of the vertical interface symmetry the initial relations $\phiAB_n=\phiBC_n$
then remains fulfilled for all times $t$.
Each of the two independent Lagrangians (\Eq{eq:Lc_AA1BB1AA1_reduced_equalpol})
turns out to be identical to that of an equivalent 1-bit interface, \Eq{eq:Lc_BB1}, 
whose center junction equals the A-line junction of the 2-bit interface, ($\CJaa, \IJaa$), and whose interface inductance $\Lhat$ likewise is determined by the A-line inductance $\Lhat_A$.
This equivalence is illustrated in Fig.~\ref{fig:interface_AA1BB1CC1}(b).

\paragraph*{Case II, opposite polarities:}
Here $\phiBC = -\phiAB$ and $\dot{\phi}^{BC} = -\dot{\phi}^{AB}$. 
The cell constraints then imply that $\phiAA=\phiCC$. 
The currents in the junctions with $\phiAA$ and $\phiCC$ are equal and their sum is 
compensated by the current in the B-line.
Again, the interface Lagrangian becomes a sum of two independent contributions,
\Eq{eq:Lc_AA1BB1AA1_reduced_oppositepol},
and each of them is identical to that of an equivalent 1-bit interface,
compare \Eq{eq:Lc_AA1BB1}.
This equivalence is shown in the schematics of Fig.~\ref{fig:interface_AA1BB1CC1}(c).
However, compared with the 1-bit interface discussed in the previous case, 
this one has center junctions both in the A- and the B-line
(cf.~Fig.~\ref{fig:interface_AA1BB1}).
In this equivalent 1-bit interface
the A-line junction is identical to the original A-line junction 
of the 2-bit interface, with ($\CJaa, \IJaa$).
In contrast, the B-line junction has only half of the original capacitance
and critical current ($\CJbb/2$,$\IJbb/2$).
As a result its junction phase $\phiBB$ is identical to that in the 2-bit interface, 
while it carries only half of the current, 
$-I = (\Phi_0/2\pi)(\CJbb/2) \ddot{\phi}^B + (\IJbb/2) \sin\phiBB$.
The sum of inductances in the equivalent 1-bit interface 
equals the sum $(\Lhat_A + 2\Lhat_B)$ of the 2-bit interface.
Furthermore, if the equivalent 1-bit interface itself has vertical symmetry,
i.e.~if $\CJaa \approx \CJbb/2$
and if the critical currents $\IJaa, \IJbb/2$ are negligible, 
then $\phiBB \approx -\phiAA$ in the equivalent 1-bit interface with 2 center junctions. 
The dynamics becomes approximately equivalent to that in a 1-bit
interface with only one interface junction, which thus has the equivalent serial capacitance $\CJaa/2$, as shown in Fig.~\ref{fig:interface_AA1BB1CC1}(d).
See App.~\ref{app:CCM_extended} for a detailed discussion 
including CC analysis of the new 1-bit interface. 
The CC analysis can also be extended to model other 2-bit gates, 
if asymmetry prevents a mapping to the two dynamically decoupled 1-bit gates.

In summary, these mappings to 1-bit circuits imply that synchronized fluxons in 2-bit circuits
can be made to undergo the forward scattering of 1-bit gates.
Moreover, as the mappings are different for equal and opposite polarities of the input fluxons, 
a 2-bit gate can be designed which deterministically
carries out the NOT or ID processes as a controlled gate.
An example for this is the controlled NOT SWAP (NSWAP) gate,
described above and in Fig.~\ref{fig:2bit_gate} for a 2-bit circuit 
(with interface parameters
$C_J^{A}/C_J = 12.2$, $C_J^{B}/C_J = 22.0$, $\CJab / C_J = 5.82$,
$I_c^{A}/I_c = 0.02, I_c^{B}/I_c = 0.01$, $\IJab / I_c = 0.53$,
$\Lhat^A/L = 0.30$, and $\Lhat^B/L = 0$).
For equal polarity fluxons, the equivalent 1-bit interface 
has, according to the mapping of Fig.~\ref{fig:interface_AA1BB1CC1}(b),
the inductance $\Lhat = 0.30 L$ and capacitances $\CJaa = 12.2 C_J$
and $\CJab = 5.82 C_J$,
while the critical current of the interface junction is negligible, $\IJaa \ll I_c$.
Comparing this to the 1-bit gate-optimization curves in Fig.~\ref{fig:optimparams__CJbb}
(with line-index $B$ instead of $A$)
it is clear that these parameters support a NOT gate, individually executed on both input fluxons.
Thus, the input bit-state pairs (0,0) and (1,1) are transformed 
to output state pairs (1,1) and (0,0), respectively.
For opposite polarity fluxons the equivalent 1-bit interface
has according to the mapping 
of Fig.~\ref{fig:interface_AA1BB1CC1}(c)
a B-junction with capacitance $\CJbb/2 = 11.0 C_J$,
while all other interface capacitances are invariant, 
$\CJaa = 12.2 C_J$ and $\CJab = 5.82 C_J$.
When we simulate the fluxon dynamics for this equivalent interface, using 
the interface Lagrangian for two center junctions, 
\Eq{eq:Lc_AA1BB1} instead of \Eq{eq:Lc_BB1} for one center junctions,
we observe forward scattering without polarity inversion, similar to Fig.~\ref{fig:4bounce}.
Because this 1-bit interface has approximate vertical symmetry,
$\CJaa \approx \CJbb/2$ and negligible $I_c^{A,B}$,
we can further compare it with an (approximately) equivalent 1-bit interface with only one center junction
of serial capacitance $\CJaa/2 \approx 6.1 C_J$,
as illustrated in Fig.~\ref{fig:interface_AA1BB1CC1}(d).
Again, from Fig.~\ref{fig:optimparams__CJbb}
it is clear that this interface supports an ID gate.
For the 1-bit interface this means that the synchronized fluxons here both perform an ID gate,
such that input state pairs (0,1) and (1,0) are preserved. 
The combination of the two input cases creates a NSWAP gate, a gate that  depends on the strong interaction of the input fields.

\section{Summary and conclusion}
\label{sec:summary}

We have proposed Reversible Fluxon Logic (RFL) as new reversible logic gate circuits. The bit states are represented by the fluxon and antifluxon state in a long Josephson junction (LJJ) which have opposite topological charge (flux polarity). 
The RFL gates are built from LJJ segments and a circuit interface between them.
The key physical feature of the gates is the resonant nonlinear dynamics induced by the incoming fluxon at the interface. 
We find through numerical circuit simulations that the fluxon is converted at the interface to and then from a localized oscillation mode, thus realizing a forward scattering of the fluxon from one LJJ to the other.  
Both the moving fluxon and the localized interface mode have a finite width on the order of the Josephson penetration length $\lambda_J$, and this 
facilitates the resonant conversion between these two excitation types. 
This mechanism therefore distinguishes RFL gates from lumped-element circuits used in SFQ-based (reversible) logic. 

Importantly, in the process of forward-scattering at the interface the fluxons can undergo conditional changes of their polarity. This effect provides the means of bit-switching in the RFL gates. 
The bit-switching is characterized by a $4\pi$-change of a central JJ in the gate interface 
(see Fig.~\ref{fig:interface}), 
contrasting the $2\pi$-switching of strongly damped JJs in SFQ digital logic.  
The dynamics in the RFL gate circuits is powered only by the inertia of incident fluxons, which distinguishes them from previous reversible digital circuits which use power from an adiabatic clock for dynamics. 
At the same time, while being a ballistic-type of reversible logic, RFL also differs from the classic billiard ball model in that the digital state is not encoded in the particle paths but their topological charge (flux polarity). Contrary to the billiard ball scattering-based gates in 2D, no path correction is therefore needed in RFL gates since given 1D paths are defined by the LJJs for the result.

As shown in Fig.~\ref{fig:sensitivity}, 
the scattering process at the interface depends on the circuit parameters and (weakly) on the input velocity of the fluxons. 
RFL gates are defined with those interface parameters which enable the desired scattering type for fluxons at a moderately high input velocity (e.g. $v=0.6 c$), with a high energy retention of the output fluxons. 

The fundamental building blocks of the logic are the 1-bit gate circuits which we design with large shunt capacitances of the three interface JJs (Fig.~\ref{fig:interface}). These capacitances absorb and later release a large fraction of the incoming fluxon's energy. This process is aided by the evanescent field that is excited around the interface over the Josephson penetration length $\lambda_J$.  Depending on the interface parameters, the new fluxon has inverted or unchanged polarity, and the two circuits therefore define the NOT and Identity (ID) gates, respectively. We note that the NOT and ID gate parameters are not unique; a NOT gate may be obtained from an ID gate by modifying the critical current or capacitance in the interface-center JJ.  For example, we showed different NOT gates in Figs.~\ref{fig:2bounce} and \ref{fig:CCM_Epot_others} related to the ID gate of Fig.~\ref{fig:4bounce}.

As there is no external power supply during the gate operation, the energy cost of an RFL gate is given simply by the energy difference between the output and the input fluxons. The fluxon output velocity and energy is calculated for various gate parameters (see Fig.~\ref{fig:optimparams__CJbb}). 
In our simulations with particular gate parameters, the output fluxons recover $>97 \%$ 
of the input fluxon energy, which is
$\Efl = 10 E_0 = 10 \Phi_0 I_c \lambda_J/(2\pi a) $ at a speed $v = 0.6c$. 

In a digital architecture consisting of many RFL gates, of course additional structures are required where fluxons are synchronized and brought back to their nominal speed. Even if fluxons are stopped in such components, the entire potential energy of fluxons could be conserved (i.e. their rest mass which is $80\%$ in our study). 
In later work \cite{WusOsb2018} we describe a circuit structure that allows one to store and launch fluxons for synchronization before entering a ballistic gate. The energy for accelerating the stored fluxons in that structure is supplied  by a clock fluxon with low energy relative to the data fluxon.

The operation time of the here presented RFL gates is only a few Josephson oscillation periods $1/\nu_J$ such that the gates are fast as well as efficient. Compared with this, the gate cycle in adiabatically-powered reversible gates  uses many oscillation periods of a JJ for the operation time in order to meet the adiabatic criteria, to conserve most of the digital state energy  $\sim I_c \Phi_0$.

A 2-bit NSWAP gate was studied as a natural extension of the 1-bit gates. 
It exhibits (see Fig.~\ref{fig:2bit_gate}) 
one of two types of dynamics, depending on whether the input fluxon polarities are the same or different. 
For all possible input polarities a dynamically equivalent 1-bit gate can be found, 
i.e. the coupled dynamics of the 2-bit gate is in each case mapped (see Fig.~\ref{fig:interface_AA1BB1CC1}) 
to that of two uncoupled 1-bit gates.
The 2-bit NSWAP was also numerically simulated (see Fig.~\ref{fig:2bit_gate_testplatform}) in a proposed experimental test platform 
(see Fig.~\ref{fig:interface}(e)) 
which features capacitive coupling to a ground plane and an imperfect fluxon launch. 
This simulation shows that the gate operation is robust even in the presence 
of non-optimized stray capacitance and launch-induced plasma waves that perturb the gate dynamics. 
Simulations of 1-bit gates were made over a range of parameters, and the output velocity shows that gates are compatible with current fabrication uncertainties (see Fig.~\ref{fig:sensitivity}).

To explain the numerically discovered phenomena of fluxon conversion to resonant excitation, followed by fluxon (or antifluxon) creation we developed and analyzed a  collective coordinate model. 
In this model we parametrize the fields in each LJJ as a superposition of fluxon and mirror antifluxon and thus reduce the many-junction dynamics to that of only two coupled coordinates. 
We solve the resulting reduced system and find quantitative agreement with the solution of the numerical circuit simulation (see Fig.~\ref{fig:CCM}).
The model describes motion of fluxons and antifluxons in the LJJs as motion in four valleys of a two-dimensional potential which may also be connected to allow scattering between valleys. 
The energy-conserving scattering process is described not by fluxons, but as particles that change between fluxon and antifluxon types smoothly in time. 
The influence of kinetic coupling and mass gradients stemming from the interface are essential to the gate dynamics since there is no external modulation of the potential. 

Reversible logic is now successfully realized in recent demonstrations of circuits that commonly share an adiabatic drive to steer the dynamics. However, we find that reversible gates are possible with an unrealized type known as ballistic reversible gates.
Our collective coordinate model for 1-bit gates describes the gate dynamics in terms of particles moving in a static potential under the influence of mass-gradient forces.  
We expect future experimental studies of these gates for scientific and technological purposes. 
We also provided estimates for energy limitations related to timing and launching a fluxon, as low as the order of $k_B T$, when used in a future possible architecture.

This provides an efficiency benefit over irreversible gates by orders of magnitude. Also, the addition of ballistic gates enhances the breadth of reversible digital logic. This may ultimately be useful to speed development in reversible digital logic similar to the way a broad set of superconducting qubit types advanced quantum reversible logic.

\section*{Acknowledgments} 
KDO and WW acknowledge useful discussions with Q.~Herr, V.~Semenov, F.~Gaitan,
and R.~Ruskov. 
WW acknowledges support from University Technical Services.

\FloatBarrier
\begin{appendix}{}

\section{Fluxon radiation in a discrete LJJ}
\label{app:discreteness}

The fluxon in an ideal continuous LJJ can move without energy loss,
as described by the soliton solution to the sine-Gordon equation.
In a discrete LJJ, as described by the Frenkel-Kontorova model \cite{BraunKivshar1998}, 
the discreteness acts as a perturbation to the Sine-Gordon dynamics. 
This perturbation causes a coupling between the fluxon and the spectrum of linear plasma waves. 
As a result, a moving fluxon excites plasma waves and loses energy in a resonant process known as Cherenkov radiation \cite{BraunKivshar1998, UstinovETAL2008}. 
For small to moderate discreteness, such as the one used in this work, 
this process is inefficient, whereas at large discreteness the energy loss due to Cherenkov radiation becomes strong \cite{PeyKru1984}.

\begin{figure}[bt]
\centering
\includegraphics[width=0.97\columnwidth]{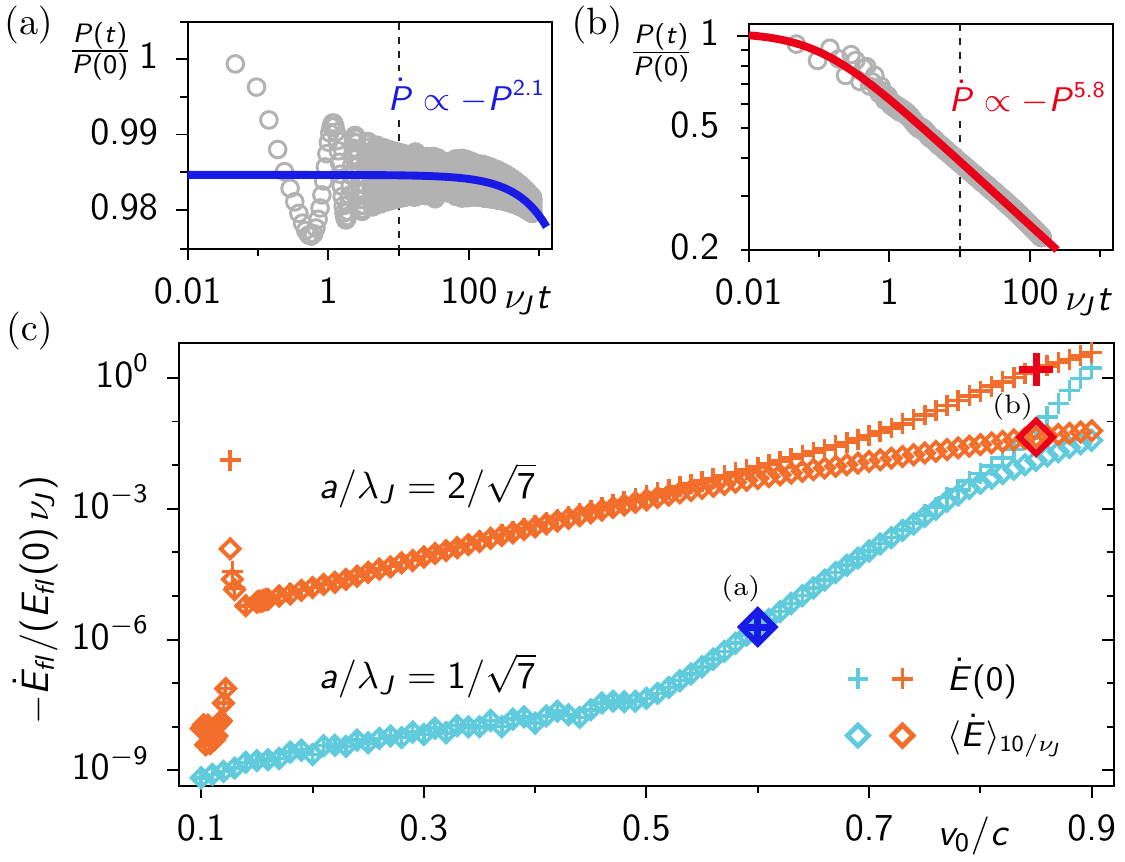}
\caption{
(a,b) Simulated momentum evolution $P(t)$ vs.~$t$ (marker) and 
fits with \Eq{eq:momentum_powerfit} (solid line), 
for fluxons initialized with (a) $v_0/c=0.6$ at $a/\lambda_J=1/\sqrt{7}$ and 
(b) with $v_0/c=0.85$ at $a/\lambda_J=2/\sqrt{7}$.
(c) Dissipation rate $\dot{E}_{\text{fl}}$ of the fluxon energy vs.~initial velocity $v_0$
for $a/\lambda=1/\sqrt{7}$ (light blue) 
and $a/\lambda=2/\sqrt{7}$ (orange).
The time-dependent dissipation rate $\dot{E}_{\text{fl}}(t)$, \Eq{eqA:dEfl_dt}, is calculated from fits of momentum $P(t)$ with \Eq{eq:momentum_powerfit}, as in examples (a,b). 
Special markers correspond to $P(t)$ shown in panels (a) and (b).
We show the initial rate $\dot{E}_{\text{fl}}(t=0)$ (plus), 
using $P(0)$ and $\dot P(0)$ from \Eq{eq:momentum_powerfit},
and the average rate $\la \dot E \ra$ over time $\nu_J t = 10$ (diamond);
this averaging interval is indicated by the dashed line in panels (a,b). 
}
\label{fig:dEfl}
\end{figure}

Here we simulate fluxon motion in our discrete LJJ to show different damping regimes.
The dissipation rate of the fluxon energy  $\dot{E}_{\text{fl}}$ is strongest 
at ultra-relativistic speeds, $v_0/c \to 1$, 
and decreases by many orders of magnitude for moderate values of $v_0$, 
see $v_0/c=0.9$ versus $0.6$ in Fig.~\ref{fig:dEfl}(c).
For fixed value of $v_0$ below the ultra-relativistic regime (such as $v_0/c=0.6$)
the dissipation rate also 
drops by orders of magnitude when the effective lattice spacing $a/\lambda_J$ is decreased. 
For example, in Fig.~\ref{fig:dEfl} we compare dissipation rates 
for our default value, $a/\lambda_J = 1/\sqrt{7}$
(light blue), with the doubled value, $a/\lambda_J = 2/\sqrt{7}$
(orange). 
At a velocity $v_0/c = 0.6$ the dissipation rate of the latter is 
$|\dot{E}_{\text{fl}}|/E_{\text{fl}} \approx 10^{-3} \omega_J$ which may not be negligible
e.g.~on a time scale of some $100 \omega_J^{-1}$. 
Compared to that the dissipation rate for the default discreteness is reduced by orders of magnitude, 
$|\dot{E}_{\text{fl}}|/E_{\text{fl}} \approx 3\cdot 10^{-7} \omega_J$,
and is negligible in the context of this study of gate times on the order of $\omega_J^{-1}$.

The dissipation rate $\dot{E}_{\text{fl}}$ presented in Fig.~\ref{fig:dEfl} is calculated
from the time-dependent fluxon momentum,
\begin{equation}
P = 8 \dot{X} (E_0/c^2)  \left( 1-(\dot{X}/c)^2 \right)^{-1/2} 
\end{equation}
where the velocity $\dot X$ follows from the fits of the moving fluxon 
according to \Eq{eq:soliton} with fit parameters $X$ and $W$.
As above, we use the characteristic energy $E_0 = (\Phi_0/2\pi) I_c \lambda_J/a$ of the continuum equation, and the characteristic momentum is $E_0/c$.
We then fit $P$ vs.~time $t$ with the function
\begin{equation}\label{eq:momentum_powerfit}
 P(t) = \frac{\bar P}{ \left( 1 + \gamma (\alpha-1) \omega_J t \left(\bar P c/E_0\right)^{\alpha-1}
 \right)^{1/(\alpha-1)} }
\end{equation}
where $\bar P, \gamma, \alpha$ are independent fit parameters.
This equation follows from a momentum-decay rate 
proportional to a power $\alpha \geq 1$  of the momentum,
\begin{equation}\label{eqA:dPdt_powerfit}
\omega_J^{-1} \dot P = -\gamma P^{\alpha}  (c/E_0)^{\alpha-1} 
\,.
\end{equation}
We find that this function describes the momentum decay well over a large range of 
damping rates, as demonstrated in Figs.~\ref{fig:dEfl}(a,b). For very small damping rates
exponential behavior is recovered, $P(t) = \bar P \exp(-\gamma \omega_J t)$,
corresponding to momentum-proportional damping $\omega_J^{-1} \dot P = -\gamma P$
from initial momentum $\bar P$.

Finally, we evaluate the energy dissipation rate,
\begin{eqnarray}\label{eqA:dEfl_dt}
 \dot{E}_{\text{fl}}(t) &=& \frac{\dot{P} P c^2}{8 E_0} 
 \left( 1 + \frac{(c P)^2}{8 E_0^2} \right)^{-1/2}
 \,,
\end{eqnarray}
following from the momentum-energy relation 
$E_{\text{fl}} = 8 E_0 \left( 1 + (c P)^2/(8 E_0^2) \right)^{1/2}$ 
and using $P(t)$ from the momentum fit, \Eq{eq:momentum_powerfit}. 
In Fig.~\ref{fig:dEfl}(c) we present both the initial dissipation rate from the fit, 
$\dot{E}_{\text{fl}}(t=0)$ (plus), as well as the dissipation rate $\la \dot E \ra$ 
averaged over a time $\omega_J t = 50$ (diamonds), 
for two discretenesses as a function of the initial velocity $v_0$.
At large $v_0/c$ the initial decay rate is much higher than the time-averaged one, 
showing that the dissipation changes rapidly during the averaging time.
This is shown in panel (c) e.g. at $v_0/c=0.85$, indicated by the (b)-markers for the discreteness $a/\lambda = 2/\sqrt{7}$, 
and the time-dependent momentum for this initial velocity and discreteness is shown in panel (b).
The initial momentum decay rate of \Eq{eqA:dPdt_powerfit} is large,
$\dot P \approx -4 (E_0 \omega_J/c)$ at $t=0$, 
and creates a large initial dissipation rate $\dot{E}_{\text{fl}}$. 
As $\dot{E}_{\text{fl}}(t)$ decays quickly in time, $\la \dot E \ra$ is much smaller in comparison. 
At lower velocity and higher discreteness the energy loss rate is much lower.
For example, at $v_0/c=0.6$ and $a/\lambda = 1/\sqrt{7}$ 
((a)-markers in panel (c) with the corresponding time-dependent momentum shown in panel (a),
$\dot P \approx -10^{-5} (E_0 \omega_J/c)$ is very small at $t=0$, and therefore both measures coincide.

The sudden reduction of the dissipation rate observed below $v_0/c \approx 0.12$ 
for the higher discreteness $a/\lambda_J = 2/\sqrt{7}$  is related to the resonant 
emission of plasma waves at specific frequencies \cite{PeyKru1984}. 
Once the fluxon velocity falls below this critical value, 
an individual solution of the resonance condition becomes inaccessible. As a result, emission into that mode is suppressed.

\section{Collective coordinate analysis}

\subsection{Interface with one center junction (1-JJ)}\label{app:CCM_simple}

\begin{figure}[bt]
\centering
\includegraphics[width=\columnwidth]{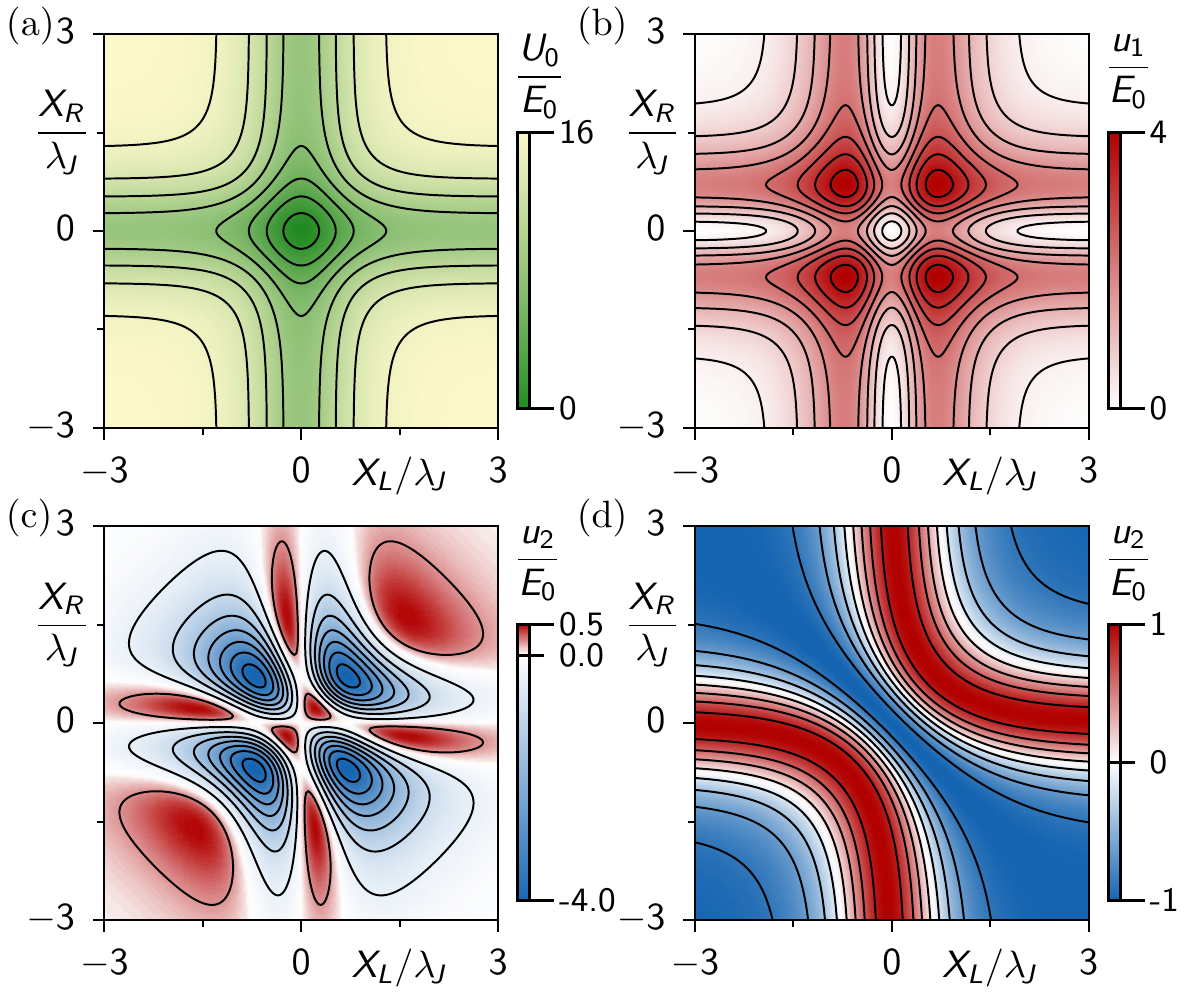}
\caption{
Contributions to CC potential $U$ vs.~$X_L, X_R$.
(a) LJJ contribution $U_0$, \Eq{eqA:U0_CCM},
and interface contributions for 1-bit interface with one center junction:
(b) $u_1$, \Eq{eqA:u1_CCM}, and (c) $u_{2}$, \Eq{eqA:u23_CCM}.  
For 1-bit interface with two center junctions $U_0$ and $u_1$ are identical to above, 
while (d) $u_{2}$ differs, \Eq{eqA:u23_CCM_AA1BB1}.
}
\label{fig:CCM_potentials}
\end{figure}

Here we present the CC analysis leading to \Eqs{eq:L_CCM}--\eqref{eq:CCM_FM}, 
for the circuit of Fig.~\ref{fig:interface}(a).
We separate the Lagrangian in \Eq{eq:Lagr_orig_discrete} as
\begin{eqnarray}\label{eq:Lagr_discrete}
 \mathcal{L} 
 &=& \tilde{\mathcal{L}} E_0 
 = \left(\tilde{\mathcal{L}}_l + \tilde{\mathcal{L}}_r + \tilde{\mathcal{L}}_I\right) E_0 
 \,,
\end{eqnarray}
where the contributions of left and right LJJ and of the interface are
\begin{eqnarray}
\label{eq:L_leftLJJ}
\tilde{\mathcal{L}}_l 
 &=& \frac{a}{\lambda_J} \sum_{n=1}^{N_l} \left[
 \frac{1}{2} \frac{\dot{\phi}_n^2}{\omega_J^2}
 + \cos\phi_n
 -\frac{(\phi_{n+1}-\phi_n)^2}{2 (a/\lambda_J)^2} 
 \right] \\
\label{eq:L_rightLJJ}
 \tilde{\mathcal{L}}_r 
&=& \frac{a}{\lambda_J} \sum_{n=N_l+1}^{N} \left[ \frac{1}{2} \frac{\dot{\phi}_n^2}{\omega_J^2} 
 + \cos\phi_n
 -\frac{(\phi_{n}-\phi_{n-1})^2}{2 (a/\lambda_J)^2} \right]\quad \\
 \label{eq:Lc_BB1}
\tilde{\mathcal{L}}_I 
 &=& \frac{a}{\lambda_J} \left\{ 
 \frac{1}{2} \frac{\CJab-C_J}{C_J \omega_J^2} \left[\dot{\phi}_L^2 + \dot{\phi}_R^2 \right]
 + \frac{1}{2} \frac{\CJbb}{C_J} \frac{(\dot{\phi}^B)^2}{\omega_J^2} \right. \\
 && - \frac{1}{2} \frac{L}{\Lhat} \frac{\lambda_J^2}{a^2} (\phiR-\phiL+\phiBB)^2 \nonumber \\
 && \hspace*{0.1cm} \left. 
 + \frac{\IJab-I_c}{I_c} \left[\cos\phiL + \cos\phiR\right] + \frac{\IJbb}{I_c} \cos\phiBB
 \right\}  \nonumber 
 \,.
\end{eqnarray}
Here we have included charging and Josephson energy for extra junctions with phases $\phi_{\nl} = \phiL$ and
$\phi_{\nr} = \phiR$ to the LJJs in \Eqs{eq:L_leftLJJ} and \eqref{eq:L_rightLJJ}.
To correct for this, the same charging energy $\propto C_J$ and Josephson energy $\propto I_c$
are subtracted in \Eq{eq:Lc_BB1} for the interface.
This allows us to replace the LJJ-sums in the continuum limit, 
$a/\lambda_J \to 0$, 
by integrals with boundaries $(-\infty, 0)$ and $(0,\infty)$.
Inserting $\phi$ and $\dot \phi$ from \Eq{eq:fluxoncombination} 
these integrations yield
\begin{eqnarray}
&&  \tilde{\mathcal{L}}_l + \tilde{\mathcal{L}}_r 
= \sum_{i=L,R} \frac{m_0(X_i)}{2} \frac{\dot{X}_i^2}{c^2} - U_0(X_L,X_R) \\
\label{eqA:U0_CCM}
&& U_0 = \sum_{i=L,R} \frac{4\lambda_J}{W} \left( 1 - \frac{2 z_i}{\sinh(2 z_i)} \right) \\
 && \hspace*{1.5cm}
 + \frac{2 W}{\lambda_J} 
 \tanh(z_i) \sech^2(z_i) \left[ 2 z_i + \sinh(2 z_i) \right] \nonumber \\
\label{eqA:m0_CCM}
&& m_0(X_i) = \frac{8\lambda_J}{W} \left(1 + \frac{2 z_i}{\sinh(2z_i)} \right)
\end{eqnarray} 
with $z_i = X_i/W$ ($i=L,R$),
and dimensionless potential $U_0(X_L,X_R)$ and masses $m_0(X_i)$.
$U_0(X_L,X_R)$ is shown in Fig.~\ref{fig:CCM_potentials}(a).

From \Eq{eq:Lagr_orig_discrete} (or \Eqs{eq:Lagr_discrete}--\eqref{eq:Lc_BB1})
we obtain the equations of motion for the interface junctions, 
\begin{eqnarray}
\label{eq:ddotphiL}
\frac{\CJab}{C_J} \frac{\ddot{\phi}_L}{\omega_J^2} 
&=& \phantom{-}\frac{\lambda_J^2 L}{a^2 \Lhat} \zeta 
- \frac{\lambda_J}{a} \frac{\phiL - \phi_{N_l-1}}{a/\lambda_J}
- \frac{\IJab}{I_c} \sin\phiL \qquad \\
\label{eq:ddotphiR}
\frac{\CJab}{C_J} \frac{\ddot{\phi}_R}{\omega_J^2}  
&=& -\frac{\lambda_J^2 L}{a^2 \Lhat} \zeta 
+ \frac{\lambda_J}{a}\frac{\phi_{N_l+2} - \phiR}{a/\lambda_J}
-\frac{\IJab}{I_c} \sin\phiR \qquad \\
\label{eq:ddotphiBB}
 \frac{\CJbb}{C_J} \frac{\ddot{\phi}^{B}}{\omega_J^2} 
 &=& -\frac{\lambda_J^2 L}{a^2 \Lhat} \zeta  - \frac{\IJbb}{I_c} \sin\phiBB
 \,,
\end{eqnarray}
where we have introduced $\zeta = (\phiR-\phiL+\phiBB)$. 
We are interested in the regime of $(L \lambda_J^2)/(\Lhat a^2) \gg \CJab/C_J$, $\lambda_J/a$, $\IJab/I_c$,
$\CJbb/C_J$, $\IJbb/I_c$.
This allows us to treat \Eqs{eq:ddotphiL}--\eqref{eq:ddotphiBB} in perturbation expansion,
with the small parameter $(\Lhat a^2)/(L \lambda_J^2)$.
The leading order contribution has $\zeta = 0$, i.e.~$\phiBB=\phiL-\phiR$, 
while $\phiL,\phiR,\phiBB$ occurring individually 
in \Eqs{eq:ddotphiL}--\eqref{eq:ddotphiBB} have finite leading-order contributions.
One can then use \Eq{eq:ddotphiBB} to determine the next-to-leading order contribution of $\zeta$,
$(L \lambda_J^2)/(\Lhat a^2) \zeta  = -\CJbb/(C_J \omega_J^2) \ddot{\phi}^{B} - (\IJbb/I_c) \sin \phi^{B}$.
Inserting this in the remaining \Eqs{eq:ddotphiL}--\eqref{eq:ddotphiR},
we obtain equations of motion for $\phiL$ and $\phiR$ in leading order.
This reduced dynamical system is still described by the Lagrangian \Eq{eq:Lagr_discrete}, 
but the interface Lagrangian now takes the form 
\begin{eqnarray}\label{eq:Lc_BB1_reduced}
&&  \tilde{\mathcal{L}}_I =
 \frac{\CJab - C_J + \CJbb}{2 C_J \omega_J^2 \lambda_J/a} \left[\dot{\phi}_L^2 + \dot{\phi}_R^2 \right]
 - \frac{\CJbb}{C_J  \lambda_J/a} \frac{\dot{\phi}_L \dot{\phi}_R}{\omega_J^2} \\
 && \hspace*{0.5cm}
 + \frac{\IJab - I_c}{I_c  \lambda_J/a} \left[\cos\phiL + \cos\phiR\right]
 + \frac{\IJbb}{I_c  \lambda_J/a} \cos(\phiL-\phiR) \nonumber 
 \,.
\end{eqnarray}

Inserting $\phi$ and $\dot \phi$ from \Eq{eq:fluxoncombination} into \Eq{eq:Lc_BB1_reduced}, where we approximate $\phi_{L} \approx \phi(x=0^{-})$  and $\phi_{R} \approx \phi(x=0^{+})$,
we calculate the interface contribution to the Lagrangian which then reads
\begin{eqnarray}
\label{eqA:L_CCM}   
&& \mkern-36mu \tilde{\mathcal{L}} 
=  \frac{m_L\dot{X}_L^2}{2c^2} + \frac{m_R\dot{X}_R^2}{2c^2} 
   + m_{LR} \frac{\dot{X}_L \dot{X}_R}{c^2} - U(X_L,X_R) \\
\label{eqA:U_CCM}   
&& \mkern-36mu U = U_0 + \frac{\IJab-I_c+\IJbb}{I_c\lambda_J/a} u_1
 + \frac{\IJbb}{I_c\lambda_J/a} u_{2}  \,, 
%
\end{eqnarray}
with coordinate-dependent masses
\begin{eqnarray}
&& m_i(X_i) = m_0(X_i) + \frac{\CJab-C_J + \CJbb}{C_J \lambda_J/a} (g_I(X_i))^2 
\end{eqnarray}
and coupling mass
\begin{eqnarray}
&& m_{LR}(X_L,X_R) = \frac{\CJbb}{C_J \lambda_J/a} g_I(X_L) g_I(X_R) \\
\label{eqA:gI_CCM}
&& g_I(X_i) = 4 \left(\lambda_J/W\right) \sech(X_i/W) 
\,,
\end{eqnarray}
compare \Eqs{eq:L_CCM}--\eqref{eq:gI_CCM}.
The interface contributions to the potential are
\begin{eqnarray}
\label{eqA:u1_CCM}
 u_1 &=& \sum_{i=L,R} 8 \sech^2(z_i) \tanh^2(z_i)  \\
\label{eqA:u23_CCM}
 u_{2} &=& - \prod_{i=L,R} \left[ 8 \sech^2(z_i) \tanh^2(z_i) \right] \\
 && + \prod_{i=L,R} \left[4 \sech(z_i) \tanh(z_i) \left( 1 - 2 \sech^2(z_i) \right) \right] \nonumber
 \,,
\end{eqnarray}
and are shown in Figs.~\ref{fig:CCM_potentials}(b) and (c).
The potential $U(X_L,X_R)$ 
is symmetric under the coordinate exchange $X_L \leftrightarrow X_R$.
The contributions $U_0$ and $u_1$ have even parity under each 
of the transformations $X_{i} \leftrightarrow - X_{i}$ ($i=L,R$),
while $u_{2}(X_L,X_R)$ has no parity symmetry.

\subsection{Interface with two center junctions (2-JJ)}\label{app:CCM_extended}

\begin{figure}[tb]
\centering
\includegraphics[width=1.0\columnwidth]{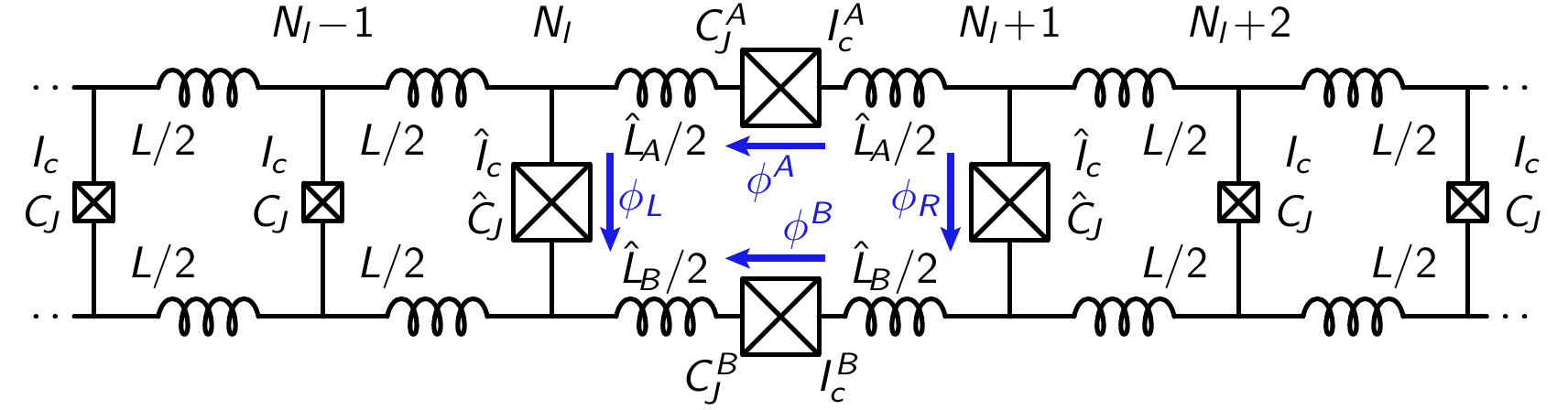}
 \caption{A 1-bit gate connecting two LJJs that has two center junctions.
 This circuit is also introduced in Fig.~\ref{fig:interface_AA1BB1CC1}(b) 
 by the equivalence to a 2-bit circuit.}
 \label{fig:interface_AA1BB1}
\end{figure}

Here we consider 
a modified 1-bit interface, shown in Fig.~\ref{fig:interface_AA1BB1}.
It is similar to that in Fig.~\ref{fig:interface}(a), 
but has a center junction in the A-line of the interface as well as in the B-line.
The phase difference $\phiAA$ over this junction provides an additional degree of freedom, and
instead of \Eq{eq:Lc_BB1} the interface Lagrangian becomes
\begin{eqnarray}
\label{eq:Lc_AA1BB1}
&& \tilde{\mathcal{L}}_I = \frac{a}{\lambda_J} \left\{ 
\frac{\CJab-C_J}{2 C_J \omega_J^2} \left[\dot{\phi}_L^2 + \dot{\phi}_R^2 \right] 
 + \frac{\CJaa(\dot{\phi}^A)^2 + \CJbb(\dot{\phi}^B)^2}{2 C_J \omega_J^2} \right. \nonumber\\
&&\, 
-\, \frac{L \lambda_J^2}{2 (\Lhat_A + \Lhat_B)a^2 } \left(\phiR-\phiL-\phiAA+\phiBB\right)^2 \\
&&\,  
\left. 
 +\, \frac{\IJab-I_c}{I_c} \left[\cos\phiL + \cos\phiR\right] + \frac{\IJbb}{I_c} \cos\phiBB 
 + \frac{\IJaa}{I_c} \cos\phiAA
 \right\}  \nonumber 
 \,,
\end{eqnarray}
where $\hat L_{A,B}$ are the interface inductances.
For brevity we refer to the interface of Fig.~\ref{fig:interface_AA1BB1}
as the 2-JJ interface, and to that of Fig.~\ref{fig:interface}(a) as the 1-JJ interface.

\paragraph*{Approximate equivalence to 1-JJ interface:}

For simplicity we also only discuss the case of vertical symmetry, 
with $\CJaa=\CJbb$ and $\IJaa=\IJbb$.
Under this condition \Eq{eq:Lc_AA1BB1} generates the equation of motion 
\begin{equation}\label{eqA:EOM_psi}
 2 \CJaa\, \ddot{\psi} = -2\IJaa \sin\psi\, \cos\left((\phiAA-\phiBB)/2\right)
\end{equation}
for the phase average
\begin{equation}
 \psi = (\phiAA + \phiBB)/2
  \,.
\end{equation}
In our simulations the interface initially has $\psi = 0$. From \Eq{eqA:EOM_psi}
it follows that $\psi = 0$ is a fixed point of the dynamics, 
such that we may set $\psi = 0$ in \Eq{eq:Lc_AA1BB1},
which then becomes
\begin{eqnarray}
\label{eq:Lc_AA1BB1_sym_psi0}
&& \tilde{\mathcal{L}}_I = \frac{a}{\lambda_J} \left\{ 
\frac{\CJab-C_J}{2 C_J \omega_J^2} \left[\dot{\phi}_L^2 + \dot{\phi}_R^2 \right] 
 + \frac{2\CJaa(\dot{\phi}^A)^2}{2 C_J \omega_J^2} \right. \nonumber\\
&&\, 
-\, \frac{L \lambda_J^2}{2 (\Lhat_A + \Lhat_B)a^2 } \left(\phiR-\phiL-2\phiAA\right)^2 \\
&&\,  
\left. 
 +\, \frac{\IJab-I_c}{I_c} \left[\cos\phiL + \cos\phiR\right]
 + \frac{2\IJaa}{I_c} \cos\phiAA
 \right\}  \nonumber 
 \,.
\end{eqnarray}
To compare this expression with the Lagrangian of the 1-JJ interface,
\Eq{eq:Lc_BB1}, we first ignore the Josephson potentials. 
The 2-JJ interface with two (identical) center junctions and values $(\Lhat_A+\Lhat_B, \phiAA, \CJaa)$
then maps exactly to the equivalent 1-JJ interface and values $(\Lhat, -\phiBB/2, 2\CJbb)$.
Further analyzing the Josephson potentials to quadratic order in the phases we find that the critical current $\IJaa$
in the 2-JJ interface corresponds to a critical current $2\IJbb$ in the 
equivalent 1-JJ interface.
That means that the scattering at the (approximately) equivalent 1-JJ and 2-JJ interfaces is very similar, provided that either $\phiBB$ remains small, 
or the critical currents are small, $\IJbb \ll \IJab, I_c$.

\paragraph*{CC analysis:}

Now we perform the CC analysis for the 2-JJ interface.
To this end we go back to \Eq{eq:Lc_AA1BB1}, 
where we again assume vertical symmetry of the interface, 
but allow for finite $\psi$.
Once again we restrict ourselves to the perturbative case of small interface inductance, 
$(\Lhat_A+\Lhat_B) \ll L \lambda_J^2/a^2$. 
To leading order we obtain the constraint $\phiAA-\phiBB = \phiR-\phiL$ for the interface cell
such that $\phiAA$ and $\phiBB$ can be expressed through $\phiR-\phiL$ and $\psi$.
With these replacements \Eq{eq:Lc_AA1BB1} reads
\begin{eqnarray}\label{eq:Lc_AA1BB1_reduced}
&& \tilde{\mathcal{L}}_I = 
\frac{\CJab-C_J+\CJaa/2}{2 C_J \omega_J^2 \lambda_J/a} \left[\dot{\phi}_L^2 + \dot{\phi}_R^2 \right]
 - \frac{\CJaa/2}{C_J \lambda_J/a} \frac{\dot{\phi}_L \dot{\phi}_R}{\omega_J^2} \nonumber\\
 && \hspace*{0.5cm}  
 + \frac{\IJab-I_c}{I_c \lambda_J/a} \left[\cos\phiL + \cos\phiR\right] \\
 && \hspace*{0.5cm}
 + \frac{\CJaa}{C_J \lambda_J/a} \frac{\dot \psi^2}{\omega_J^2}  
 + \frac{2 \IJaa}{I_c \lambda_J/a} \cos\psi \cos\left(\frac{\phiL-\phiR}{2}\right)
 \nonumber 
 \,.
\end{eqnarray}
From \Eq{eq:Lagr_discrete} together with \eqref{eq:Lc_AA1BB1_reduced} 
the CC Lagrangian is derived (similar to Sec.~\ref{app:CCM_simple}),
\begin{eqnarray}
&& \tilde{\mathcal{L}} 
= \frac{m_L}{2} \frac{\dot{X}_L^2}{c^2} + \frac{m_R}{2} \frac{\dot{X}_R^2}{c^2} 
   + m_{LR} \frac{\dot{X}_L \dot{X}_R}{\omega_J^2} \\
&&\hspace*{0.8cm}
   + \frac{m_\psi}{2} \dot \psi^2 - U(X_L,X_R) \nonumber\\
\label{eqA:U_CCM_AA1BB1}
&& U = U_0 + \frac{\IJab-I_c}{I_c\lambda_J/a} u_1
 + \frac{2 \IJaa}{I_c\lambda_J/a} \cos(\psi) \, u_{2} \\
&& m_i = m_0(X_i) + \frac{\CJab - C_J + \CJaa/2}{C_J \lambda_J/a} (g_I(X_i))^2 \\
&& m_{LR} = \frac{\CJaa/2}{C_J \lambda_J/a} g_I(X_L) g_I(X_R) \\
&& m_\psi = \frac{2\CJaa}{C_J \lambda_J/a}
\end{eqnarray}
with $U_0$, $u_1$, $m_0$ and $g_I$ identical to expressions for the 1-JJ interface 
(\Eqs{eqA:U0_CCM}, \eqref{eqA:u1_CCM}, \eqref{eqA:m0_CCM}, and \eqref{eqA:gI_CCM}). 
The $\IJbb$-proportional contribution to the interface potential is
\begin{eqnarray}
\label{eqA:u23_CCM_AA1BB1}
 u_{2} &=& - \prod_{i=L,R} \left[\sech^2(z_i) - \tanh^2(z_i)\right]  \\
 && + 4 \prod_{i=L,R} \sech(z_i) \tanh(z_i) \nonumber
 \,,
\end{eqnarray}
and is shown in Fig.~\ref{fig:CCM_potentials}(d).
It differs qualitatively from the $u_{2}$ for the 1-JJ interface,
which is given in \Eq{eqA:u23_CCM} and Fig.~\ref{fig:CCM_potentials}(c). 

The form of the reduced dynamical system for $X_L,X_R$ remains invariant, 
\Eqs{eq:EOM_CCM}--\eqref{eq:CCM_FM},
but here an additional equation of motion exists for $\psi$,
\begin{equation}
 \ddot \psi = -\omega_J^2 \frac{\partial U}{\partial \psi} 
 = \frac{2 \IJbb}{I_c\lambda_J/a} \sin(\psi) \,u_{2} 
 \,.
\end{equation}
Again, starting from the fixed point, $\psi = 0$, the CC dynamics can be compared with 
that of the 1-JJ interface. To this end we substitute
parameters as above for the (approximately) equivalent 1-JJ interface.
With this substitution most CC quantities become equal to those 
in \Eqs{eqA:L_CCM}--\eqref{eqA:u1_CCM}. 
One exceptions is the different form of the interface potential $u_{2}$, as mentioned above.
Another exception is the prefactor of the interface potential $u_1$. While in the 1-JJ interface $u_1$ has a weighting $\propto \IJbb$, \Eq{eqA:U_CCM},
no equivalent weighting $\propto \IJaa$ appears in \Eq{eqA:U_CCM_AA1BB1}.
In this study we focus on the case $\IJbb \ll I_c$ where these differences in the potential $U$
are negligible, and the fluxon scattering at the (approximately) equivalent 
1- and 2-JJ interface are therefore very similar.

\section{Mapping 2-bit gate to 1-bit gates}\label{app:2bitinterface}

Here we discuss the Lagrangian for the interface which connects two LJJs from each side, 
Fig.~\ref{fig:interface_AA1BB1CC1}(a).
The interface is assumed to be both left--right and vertically symmetric.
Using the notation $\phiAB$ and $\phiBC$ for the phases across the upper and lower LJJs, respectively,
the interface Lagrangian $\tilde{\mathcal{L}}_I$ is,
\begin{eqnarray}\label{eq:Lc_AA1BB1AA1}
 &&\tilde{\mathcal{L}}_I = \frac{a}{\lambda_J} \left\{ 
 \frac{1}{2} \frac{\CJab-C_J}{C_J \omega_J^2} \sum_s 
 \left[(\dot{\phi}_L^s)^2 + (\dot{\phi}_R^s)^2 \right] 
  + \frac{\CJbb}{2C_J} \frac{(\dot{\phi}^B)^2}{\omega_J^2}
 \right. \nonumber\\
 && \hspace*{0.6cm}
 + \frac{\CJaa}{2C_J} \frac{(\dot{\phi}^A)^2 + (\dot{\phi}^C)^2}{\omega_J^2} 
 + \frac{\IJaa}{I_c} \left[ \cos\phiAA + \cos\phiCC \right] 
\nonumber \\
 && \hspace*{0.6cm}
 - \left(\iflq\right)^2 \frac{L \lambda_J^2}{2 a^2} 
 \left[\Lhat^A (\hat I^A)^2 + \Lhat^B (\hat I^B)^2 + \Lhat^A (\hat I^C)^2 \right]  \nonumber\\
 && \hspace*{0.6cm} \left. 
 + \frac{\IJab-I_c}{I_c} \sum_s \left[\cos\phiL^s + \cos\phiR^s \right] + \frac{\IJbb}{I_c} \cos\phiBB  
 \right\}  
\end{eqnarray}
($s = AB, BC$).
Without DC bias from the left to the right of the gate, 
the currents in the interface lines fulfill $\hat I^A + \hat I^B + \hat I^C = 0$,
and
\begin{eqnarray}
&& \hat I^A = \frac{\Phi_0/(2\pi)}{\Lhat^A (\Lhat^A + 2\Lhat^B)} 
  \left[ 
   - (\Lhat^A + \Lhat^B) \phiAA + \Lhat^A \phiBB + \Lhat^B \phiCC  \right. \nonumber\\
&& \left. \hspace*{0.7cm}
+ (\Lhat^A + \Lhat^B) (\phiR^{AB} - \phiL^{AB}) + \Lhat^B (\phiR^{BC} - \phiL^{BC}) \right] \\
&& \hat I^C = \frac{\Phi_0/(2\pi)}{\Lhat^A (\Lhat^A + 2\Lhat^B)} 
  \left[ \Lhat^B \phiAA + \Lhat^A \phiBB  - (\Lhat^A + \Lhat^B) \phiCC  \right. \nonumber\\
&& \left. \hspace*{0.7cm}
-\Lhat^B (\phiR^{AB} - \phiL^{AB}) - (\Lhat^A + \Lhat^B) (\phiR^{BC} - \phiL^{BC}) \right]
\;.
\end{eqnarray}

Similar to the 1-bit interfaces we consider the limit of small interface inductances $\Lhat^{A,B}$,
which here results in the interface-cell constraints $\phiAA-\phiBB = \phiAB_R - \phiAB_L$
and $\phiBB-\phiCC = \phiBC_R - \phiBC_L$. 
The 3rd line in \Eq{eq:Lc_AA1BB1AA1} then becomes negligible. 

We discuss the cases of initially equivalent fields in the upper and lower LJJ,
$\phiBC_n = \pm \phiAB_n$.
Coupling between $\phiAB$ and $\phiBC$ occurs only within the interface.
We start with the assumption that 
$\phiBC = \phiAB$ or $\phiBC = -\phiAB$ remains fulfilled 
throughout the evolution. 
This will be confirmed below by the effective decoupling of the upper and lower 
interface cell imposed by the symmetry.

\paragraph*{Case I, $\phiBC = \phiAB$:} 
The symmetry together with the above cell constraints give $\phiAA-2\phiBB+\phiCC = 0$.
The current on the B-line cancels, $\hat I^B = 0$.
For symmetry reasons, $\phiCC = -\phiAA$, and thus $\phiBB = 0$.
Therefore the interface Lagrangian, \Eq{eq:Lc_AA1BB1AA1},
is a sum of two independent contributions,
\begin{equation}
 \tilde{\mathcal{L}}_I = \tilde{\mathcal{L}}^{AB}_I + \tilde{\mathcal{L}}^{BC}_I
 \,, 
\end{equation}
with
\begin{eqnarray}
&&\tilde{\mathcal{L}}^{AB}_I = 
 \frac{a}{\lambda_J} \left\{\!
  \frac{\CJab-C_J}{2 C_J \omega_J^2} 
  \left[ (\dot{\phi}_L^{AB})^2 + (\dot{\phi}_R^{AB})^2 \right] 
 + \frac{\CJaa}{2C_J} \frac{(\dot{\phi}^A)^2}{\omega_J^2} \right. \nonumber\\
\label{eq:Lc_AA1BB1AA1_reduced_equalpol}
 && \hspace*{0.5cm}\left. 
 + \frac{\IJab-I_c}{I_c} \left[ \cos\phiL^{AB} + \cos\phiR^{AB} \right]
 + \frac{\IJaa}{I_c} \cos\phi^A \right\}  
,
\end{eqnarray}
and $\tilde{\mathcal{L}}^{BC}_I$ given by the same expression but with the substitutions 
$AB \to BC$, $\phiAB \to \phiBC$ and $\phiAA \to -\phiCC$.
Since there is no coupling term, $\phiAB$ and $\phiBC$ effectively evolve independently. 
Because of vertical symmetry of the interface 
the initial relation $\phiAB(0)=\phiBC(0)$
remains fulfilled for all times, i.e.~the fields remain synchronized.
Each of the Lagrangians $\tilde{\mathcal{L}}_I^s$
is identical to that of an equivalent 1-bit interface with one center junction,
\Eq{eq:Lc_BB1}, 
which has characteristic values ($\CJaa, \IJaa$),
as indicated in Fig.~\ref{fig:interface_AA1BB1CC1}(b).
Note that the equivalence to the 1-bit interface holds also for finite interface 
inductances $\Lhat^A, \Lhat^B$, which we have neglected here for simplicity.
The inductance $\Lhat$ of the equivalent 1-bit interface is then given by $\Lhat_A$.

\paragraph*{Case II, $\phiBC = -\phiAB$:} 
Here  the cell constraints imply that $\phiAA=\phiCC$.
Again, \Eq{eq:Lc_AA1BB1AA1} 
is a sum of two independent contributions, with
\begin{eqnarray}
&&\tilde{\mathcal{L}}^{AB}_I = 
 \frac{a}{\lambda_J} \left\{ 
  \frac{1}{2} \frac{\CJab-C_J}{C_J \omega_J^2} 
  \left[(\dot{\phi}_L^{AB})^2 + (\dot{\phi}_R^{AB})^2 \right] 
  + \frac{\CJaa}{2C_J} \frac{(\dot{\phi}^A)^2}{\omega_J^2} \right. \nonumber\\
\label{eq:Lc_AA1BB1AA1_reduced_oppositepol} 
 && \hspace*{1.0cm}
 + \frac{\CJbb}{4C_J} \frac{(\dot{\phi}^B)^2}{\omega_J^2} + \frac{\IJbb}{2 I_c} \cos\phiBB \\
 && \hspace*{1.0cm}\left. 
 + \frac{\IJab-I_c}{I_c} \left[\cos\phiL^{AB} + \cos\phiR^{AB} \right]
 + \frac{\IJaa}{I_c} \cos\phi^A 
 \right\}  \nonumber
\end{eqnarray}
where $\phiBB = \phiAA - \phiAB_R + \phiAB_L$.
The same expression defines $\tilde{\mathcal{L}}^{BC}_I$, 
but with the substitutions $AB \to BC$, $\phiAB \to -\phiBC$ and $\phiAA \to \phiCC$.
%
Each $\tilde{\mathcal{L}}_I^s$
is identical to the Lagrangian of an equivalent 1-bit interface with two center junctions,
\Eq{eq:Lc_AA1BB1}, as indicated in Fig.~\ref{fig:interface_AA1BB1CC1}(c).
The equivalent center junctions have characteristic values $(\CJaa, \IJaa)$
and $(\CJbb/2, \IJbb/2)$, respectively. Also the sum of interface inductances in the equivalent 1-bit 2-JJ interface, $\Lhat_A + \Lhat_B$, equals $\Lhat_A + 2\Lhat_B$ from the 2-bit interface.

If this equivalent 1-bit 2-JJ interface, moreover, has vertical symmetry,
i.e.~if $\CJaa=\CJbb/2, \IJaa=\IJbb/2$, then we can approximately map further to a
1-bit 1-JJ interface, as discussed in App.~\ref{app:CCM_extended}.
This approximately equivalent interface has center-junction values
$(\CJaa/2, \IJaa/2)$ and center inductance $\Lhat$ of $\Lhat_A + 2\Lhat_B$.

\end{appendix}


\end{document}